\documentclass[superscriptaddress,floatfix,preprintnumbers,twocolumn]{revtex4-1}

\usepackage{amsmath}
\usepackage{amssymb}
\usepackage[hidelinks]{hyperref}
\usepackage{graphicx}

\newcommand{\imu}{\text{i}}

\newcommand{\commentOut}[1]{}
\usepackage{empheq}

\usepackage{times}

\begin{document}

\def\tensor#1{\stackrel{\:\!\leftrightarrow}{\bf #1}}

\newbox\akzentbox
\def\stdyad#1{\setbox\akzentbox=\hbox{\lower1.5pt\hbox{$\scriptscriptstyle
\kern#1\leftrightarrow$}}\ht\akzentbox=-1.5pt\dp\akzentbox=0pt\box\akzentbox}
\def\dyad#1{\buildrel\stdyad{.00em}\over{#1}}
\def\G{{\bar{\bf G}}}

\title{Optomechanical sideband asymmetry explained by stochastic electrodynamics}

\author{L.~Novotny}
\homepage{http://www.photonics.ethz.ch}
\affiliation{Photonics Laboratory, ETH Z\"urich, 8093 Z\"urich, Switzerland}
\affiliation{Quantum Center, ETH Z\"urich, 8093 Z\"urich, Switzerland}

\author{M.~Frimmer}
\affiliation{Photonics Laboratory, ETH Z\"urich, 8093 Z\"urich, Switzerland}

\author{A.~Militaru}
\affiliation{Photonics Laboratory, ETH Z\"urich, 8093 Z\"urich, Switzerland}

\author{A.~Norrman}
\affiliation{Photonics Laboratory, ETH Z\"urich, 8093 Z\"urich, Switzerland}
\affiliation{Insitute of Photonics, University of Eastern Finland, P.O. Box 111, FI-80101 Joensuu, Finland}

\author{O.~Romero-Isart}
\affiliation{Institute for Quantum Optics and Quantum Information of the Austrian Academy of Sciences, A-6020 Innsbruck, Austria}
\affiliation{Institute for Theoretical Physics, University of Innsbruck, A-6020 Innsbruck, Austria}

\author{P.~Maurer}
\affiliation{Institute for Quantum Optics and Quantum Information of the Austrian Academy of Sciences, A-6020 Innsbruck, Austria}
\affiliation{Institute for Theoretical Physics, University of Innsbruck, A-6020 Innsbruck, Austria}

\begin{abstract}
Within the framework of stochastic electrodynamics we derive the noise spectrum of a laser beam reflected from a suspended mirror. The electromagnetic field follows Maxwell's equations and is described by a deterministic part that accounts for the laser field and a stochastic part that accounts for thermal and zero-point background fluctuations.
Likewise, the mirror motion satisfies Newton's equation of motion and is composed of  deterministic and stochastic parts, similar to a Langevin equation. We consider a photodetector that records the power of the field reflected from the mirror interfering with a frequency-shifted reference beam (heterodyne interferometry). We theoretically show that the power spectral density of the photodetector signal is composed of four parts: (i) a deterministic term with beat notes, (ii) shot noise, (iii) the actual heterodyne signal of the mirror motion and (iv) a cross term resulting from the correlation between measurement noise (shot noise) and backaction noise (radiation pressure shot noise). The latter gives rise to the Raman sideband asymmetry observed with ultracold atoms, cavity optomechanics and with levitated nanoparticles. Our classical theory fully reproduces experimental observations and agrees with the results obtained by a quantum theoretical treatment.
\end{abstract}

\maketitle

\section{Introduction}
\begin{figure*}[!t]
\begin{center}
\vspace{1.0ex}
\includegraphics[width=0.75\textwidth]{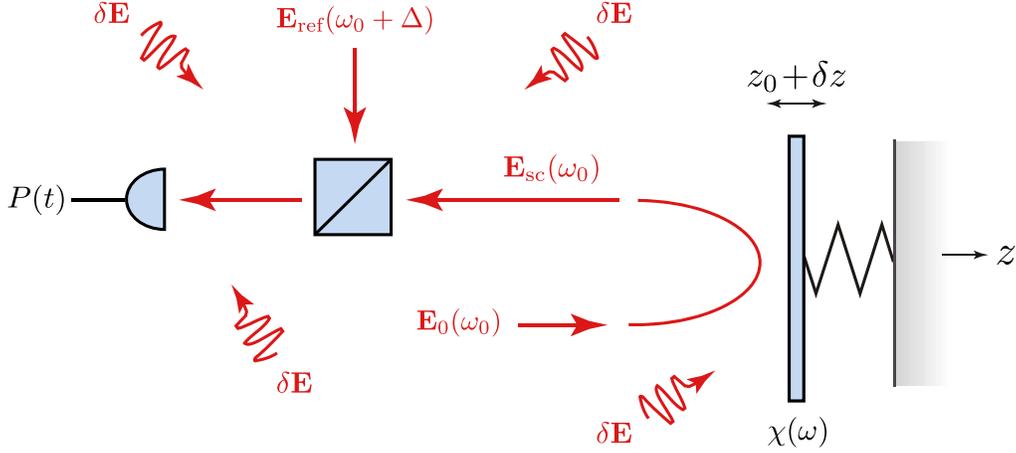}
\vspace{-1.0ex}
\end{center}
\caption{A field ${\bf E}_0$ of frequency $\omega_0$ is reflected from a suspended mirror. The reflected field ${\bf E}_{\rm sc}$ is combined with a reference field ${\bf E}_{\rm ref}$ of frequency $\omega_0+\Delta$ and directed on a photodetector that measures the power $P(t)$. The mirror is characterized by the susceptibility $\chi(\omega)$, which transduces the radiation pressure force ${\bf F}$ into a mirror displacement $z_0$. A stochastic background field $\delta {\bf E}$, originating from thermal and zero-point fluctuations, is superimposed to all fields. Similarly, a stochastic displacement $\delta z$ accounts for  thermal and zero-point fluctuations of the mirror.
}
\label{heterodyneillustration}
\end{figure*}

In Raman scattering a photon scatters from a target and thereby looses or gains a quantum of vibrational energy of the target~\cite{raman28,placzek34}. The two processes are called Stokes and anti-Stokes Raman scattering, respectively. Because of the existence of a vibrational ground state the Stokes process is more probable, by a factor given by the Bose-Einstein distribution. As a consequence, the Stokes and anti-Stokes peaks in a Raman scattering spectrum are unequal in height and their ratio can serve as a thermometer to determine the local temperature~\cite{kip90,wehrmeyer96}. A similar effect is observed in heterodyne spectroscopy where the light scattered from a target is interfered with a frequency-shifted reference beam (c.f. Fig.~\ref{heterodyneillustration}). Here the spectrum consists of two peaks centered around the modulation frequency of the reference beam. As in Raman scattering the two peaks are unequal in height, an effect termed {\em Raman sideband asymmetry}. The sideband asymmetry has been measured with several systems, including atoms~\cite{monroe95,brahms12}, spin systems~\cite{kohler17},  mechanical oscillators~\cite{safavinaeini12,weinstein14,underwood15,purdy15,sudhir17} and levitated particles~\cite{tebbenjohanns20}. Its practical use for thermometry has been discussed by Purdy et al.~\cite{purdy15}.

Despite the similarity of Raman scattering  and heterodyne spectroscopy there is an important difference. In Raman scattering one directly measures the  scattered light whereas in heterodyne spectroscopy one measures an interferometric signal. While the asymmetry in Raman scattering has a clear explanation the origin of the sideband asymmetry in heterodyne measurements is still being debated~\cite{khalili12,weinstein14,borkje16,sudhir17,machado21}. Does the sideband asymmetry originate from quantum fluctuations of the field, the target, or both? Khalili et al. interpret the asymmetry as arising from the quantum coherence between the mechanical oscillator (target) and the detector, which builds up during the measurement process and gives rise to correlations between measurement noise and back-action noise~\cite{khalili12}. Weinstein et al. point out that the interpretation of the sideband asymmetry depends on the measurement scheme and that the imbalance can be attributed to the quantum fluctuations of either the mechanical mode or  the electromagnetic field~\cite{weinstein14}. B{\o}rkje substantiates this viewpoint and finds that the interpretation depends  on the choice of the photodetector model, that is, whether the detector measures symmetrized expectation values or normal- and time-ordered expectation values~\cite{borkje16}. For a detector that measures symmetric, non-ordered expectation values, B{\o}rkje finds that the sideband asymmetry 
is the result of interference between quantum noise in the electromagnetic field and the position of the mechanical oscillator, whereas for a detector that measures normal- and time-ordered expectation values the sideband asymmetry is a direct reflection of the quantum asymmetry of the position noise of the mechanical oscillator. In a recent study, Machado and Blanter point out that in cavity optomechanics the mechanical system is embedded in an optical cavity, which further complicates the interpretation~\cite{machado21}. The objective of this paper is to show that the sideband asymmetry can be quantitatively derived without the use of quantum mechanics. We use stochastic electrodynamics, a classical theory that accounts for zero-point fluctuations~\cite{marshall63,boyer75a,boyer11}, to derive the sideband asymmetry for a simple cavity-free model. Our results reproduce previous results and provide an intuitive interpretation.

This article is structured as follows: Following this introduction we provide a short review of stochastic electrodynamics (section~\ref{stochel}). We then use this framework to calculate the shot noise measured by a photodetector  (section~\ref{shotnoise}). This calculation outlines the main theoretical steps and serves as a reference for later calculations. In section~\ref{suspmirror} we then tackle the problem of a plane wave that is reflected from a suspended mirror and then superimposed onto a frequency-shifted reference beam. The reflected plane wave and the frequency-shifted reference beam are directed onto a photodetector, which measures the optical power. Based on the optical power we evaluate the power spectral density (PSD). In section~\ref{discussionpsd} we analyze the different terms contributing to the PSD and discuss the results. This is followed by section~\ref{conclusionspsd} in which we summarize our main findings.  
In the appendix, we provide further details on our derivation, discuss the case of homodyne detection and, as a reference, provide a quantum mechanical treatment of the same problem.

\section{Stochastic Electrodynamics \label{stochel}}
In stochastic electrodynamics~\cite{marshall63,boyer75a,boyer11,novotny12} one assumes that all excitations (matter and field) are composed of a deterministic part and a stochastic part. Accordingly,  the electric field at position ${\bf r}$ and time $t$ must be expressed as ${\bf E}({\bf r},t) = {\bf E}_0({\bf r},t) + \delta{\bf E}({\bf r},t)$,
where ${\bf E}_0$ denotes the deterministic part and $\delta{\bf E}$ denotes the stochastic, statistically stationary fluctuations.  We assume the electromagnetic field to be in thermal equilibrium at a temperature $T$. This equilibrium condition dictates a relation between the fluctuations and loss channels of the system, whereupon the cross-spectral density of the stochastic field at points ${\bf r}$ and ${\bf r}'$ is governed by the 
 fluctuation dissipation theorem (FDT)~\cite{callen51,agarwal75,rytov88}
\begin{multline}
S_{  E_j  E_k}({\bf r}, {\bf r}',\omega)\equiv\frac{1}{2\pi} \int_{\mathbb{R}} d \tau  \big\langle \delta E_j({\bf r},t)\delta E_k({\bf r}',t+\tau)\big\rangle\,  {\rm e}^{\imu\omega \tau}
 \\
= \frac{\mu_0  \omega}{\pi} \left[
\frac{\hbar \omega}{2} \coth \left(\frac{\hbar\omega}{2 k_B T}\right)
\right]
{\rm Im}\left[G_{jk}({\bf r}, {\bf r}', \omega)\right]  ,
   \label{coh02de2}
\end{multline}
where $\delta E_j$ is the component along the Cartesian unit vector along the $j-$axis, $\tau=t-t'$ the time separation (following from stationarity), $\hbar$ the reduced Planck constant, $k_B$ the Boltzmann constant, and $\mu_0$ the vacuum permeability. The average $\langle \cdot \rangle$ denotes a statistical average over many realizations (which coincides with the time average as the field fluctuations are ergodic), the expression in square brackets is the average energy of a mode with angular frequency $\omega$ that tends to $\hbar|\omega|/2$ in the zero temperature limit, and $G_{jk}$ is the Cartesian element of the equilibrium Green's tensor $\G$ that consists equally of an outgoing  ($\G^+$) and an incoming part ($\G^-$) and is zero at infinity. This superposition ensures that all charges are in equilibrium with the radiation field~\cite{eckhardt82}. According to Eq.~(\ref{coh02de2}), fluctuations of the field are balanced by the dissipation represented by the imaginary part of $\G$.

In 1969 T. H. Boyer showed that the energy density of a fluctuating field at zero temperature must be proportional to the angular frequency $\omega$ (per mode), otherwise the spectrum would not be Lorentz invariant~\cite{boyer69}. Planck's constant is then introduced as a scale parameter. Expressing the fluctuating vacuum field as
\begin{multline}
\delta{\bf E}({\bf r},t)\,\equiv\, \sum_{\sigma=1}^2 \int_{\mathbb{R}^3} \!d{\bf k}  \;|\delta E(\omega_{\bf k})|\,
{\bf n}_{\sigma}({\bf k})\\[-1ex]
  \times \cos[{\bf k}\!\cdot\!{\bf r} -\omega_{\bf k} t + \theta_{\sigma}({\bf k})] ,
\label{vacfieldboyerzz}
\end{multline}
where $\omega_{\bf k} =c |{\bf k}|$ is the angular frequency that corresponds to the wave vector ${\bf k}$, with $c$ being the vacuum speed of light, $\sigma$ denotes the two transverse polarizations, $\theta_{\sigma}({\bf k})$ is a random phase and $\delta E(\omega_{\bf k})$ is the spectrum of the fluctuating field, the two-point correlation can be shown to read~\cite{boyer75b}
\begin{multline}
\big\langle \delta E_i({\bf r},t)\delta E_j({\bf r}',t+\tau)\big\rangle
 =
\int_{\mathbb{R}^3} \!d{\bf k} \;
\frac{\hbar\omega_{\bf k}}{16 \pi^3 \varepsilon_0}
(\delta_{ij} - k_i k_j / k^2) \qquad   \\
\times \cos[{\bf k}\!\cdot\!({\bf r}-{\bf r}')+ \omega_{\bf k}\tau], \label{twopointcorr}
 \end{multline}
where $\varepsilon_0$ denotes the vacuum permittivity,
$k_i$ the wave vector component along the $i$-axis, $k=|{\bf k}|$ the wave number, and $\delta_{ij}$ the Kronecker delta. It can be shown that this result is in accordance with the FDT in Eq.(\ref{coh02de2}) in the zero-temperature limit.

In analogy to the electric field we also decompose the mechanical degree of freedom $z(t)$, 
i.e. the position of the suspended mirror along the $z$-axis,
 into a deterministic and a stochastic part $z(t) = z_0(t) + \delta z_\text{rp}(t) +\delta z_\text{th}(t)$, where $z_0$ and $\delta z_\text{rp}$ denote the response to the deterministic and stochastic parts of the electromagnetic field, respectively, and $\delta z_\text{th}$ denotes the position fluctuations due to the coupling to an effective thermal bath, e.g. residual gas at temperature $T$. The corresponding FDT is~\cite{callen51}
\begin{multline}
S^\text{th}_{ z z}(\omega)= \frac{1}{2\pi} \int_{\mathbb{R}}d\tau   \big\langle \delta z_\text{th}(t)\,\delta z_\text{th}(t+\tau)\big\rangle\, {\rm e}^{\imu\omega \tau}  \\ 
=  \frac{1}{ \pi  \omega}
\left[
\frac{\hbar\Omega_0}{2} \coth\!\left(\frac{\hbar\Omega_0}{2 k_B T}\right)
\right]\,
{\rm Im}[\chi(\omega)] 
   \label{coh02de3q}, 
\end{multline}
or equivalently $S^\text{th}_{ z z}=|\chi|^2 S^\text{th}_{ F F}$ with the force spectral density
\begin{equation}\label{fluctuations}
    S^\text{th}_{ F F}= \frac{\gamma m}{ \pi }
\left[
\frac{\hbar\Omega_0}{2} \coth \left(\frac{\hbar\Omega_0}{2 k_B T}\right)
\right].
\end{equation}
Here, the fluctuations $\delta z_\text{th}$  are balanced by the losses expressed by the imaginary part of the susceptibility that reads
\begin{eqnarray}
\chi(\omega)= [m(\Omega_{0}^2\!-\omega^2-\imu \gamma \omega)]^{-1},
\label{chi}
\end{eqnarray}
where $m$ is the mass of the suspended mirror, $\Omega_0$ its natural mechanical frequency and $\gamma$ its damping.  In section~\ref{suspmirror} we will study the interaction of the stochastic field with the suspended mirror and make use of both the FDT for the electromagnetic field (\ref{coh02de2}) and the FDT of the mechanical motion (\ref{coh02de3q}).

The interplay of the fluctuating electromagnetic and mechanical degrees of freedom forms the framework of stochastic electrodynamics. In the past, it has been applied to  various phenomena, including van der Waals and Casimir forces~\cite{henkel02,novotny12}, blackbody radiation~\cite{boyer69}, heat transfer~\cite{mulet01}, the ground state of the hydrogen atom and the absence of atomic collapse~\cite{cole03}, and vacuum friction~\cite{zurita04}.  \\

\section{Photodetector Shot Noise \label{shotnoise}}
\begin{figure}[!b]
\begin{center}
\includegraphics[width=0.45\textwidth]{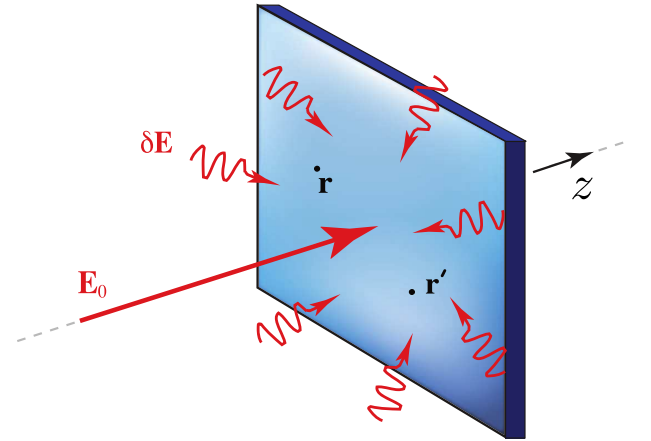}
\end{center}
\vspace{-1em}\caption{A photodetector with area $A$ is irradiated by a deterministic laser field ${\bf E}_0({\bf r},t)$ and a fluctuating vacuum field $\delta{\bf E}({\bf r},t)$. Interference between the two gives rise to shot noise.}
\label{figfluctuatingfields}
\end{figure}

Here we derive the shot noise measured by a photodetector. We consider a photodetector that renders the optical power (c.f. Fig~\ref{figfluctuatingfields})
\begin{equation}
P(t) = \int_A da\;  I(x,y,t)   ,
\label{detsigy}
 \end{equation}
where $I(x,y,t)$ is the intensity in the detector plane ($z=0$) and $A$ is the detector area, which is assumed to be much larger than any relevant wavelength. Taking into account the finite response time ${\tau_\text{det}}$ of the photodetector, the intensity is defined as \begin{equation}
I({\bf r},t) \,\equiv\, \varepsilon_0 c\,   {\bf E}({\bf r},t)\cdot{\bf E}({\bf r},t)|_{\tau_\text{det}},
\label{intdefmy}
 \end{equation}
where $|_{{\tau_\text{det}}}$ specifies that only frequencies $\omega/(2\pi)<{\tau^{-1}_\text{det}}$ are recorded. In the following we take the finite response time into account by averaging the intensity in the temporal domain and suppressing any frequency components $\omega/(2\pi)>{\tau_\text{det}^{-1}}$ in the spectral domain.

We are interested in the power spectral density (PSD) of the detector signal (\ref{detsigy}), defined as
  \begin{multline}
S_{PP}(\omega)  \equiv \frac{1}{2\pi}  \int_\mathbb{R} d\tau
\left\langle P(t) P(t+\tau)\right\rangle {\rm e}^{\imu \omega \tau}   \\
=
\frac{1}{2\pi} \int_A da \int_{A} da'   \int_\mathbb{R}  d\tau  \left\langle I({\bf r},t)\, I({\bf r}',t+\tau) \right\rangle {\rm e}^{\imu \omega \tau}  , \label{sppwm2y} 
  \end{multline}
with $\left\langle I({\bf r},t)\, I({\bf r}',t+\tau) \right\rangle$ being the intensity correlation function.

The field ${\bf E}({\bf r},t)$ incident on the detector is the sum of a deterministic monochromatic field at frequency $\omega_0= c\:\! k_0$ 
  \begin{eqnarray}
{\bf E}_0({\bf r},t) = E_0 \cos(k_0 z - \omega_0 t) \,{\bf n}_{x}, \label{sppwm2ydet} 
  \end{eqnarray}
polarized along the $x-$axis and a fluctuating vacuum field with zero mean $\langle \delta{\bf E} \rangle =0$. Neglecting terms second order in fluctuations the intensity (\ref{intdefmy}) can be represented as
\begin{align}
I({\bf r},t) &\,\simeq\,  \varepsilon_0 c  |{\bf E}_0({\bf r},t)|^2
\,+\,
2 \varepsilon_0 c \, {\bf E}_0({\bf r},t)\cdot\delta{\bf E}({\bf r},t)\\[0.5ex]
& \,\equiv\,  I_0
+ \delta I({\bf r},t) , 
\label{intdefmyd} 
 \end{align}
where $\langle \delta I \rangle = 0$ and $I_0\equiv \varepsilon_0 c E_0^2/2$. The intensity correlation function can then be written as
  \begin{eqnarray}
\left\langle I({\bf r},t)\, I({\bf r}',t+\tau) \right\rangle \,=\,
 I_0^2 +
\left\langle \delta I({\bf r},t)\, \delta I({\bf r}',t+\tau)\right\rangle,\qquad
\label{intcorr01}
 \end{eqnarray}
where
 \begin{multline}
\left\langle \delta I({\bf r},t)\, \delta I({\bf r}',t+\tau) \right\rangle \,=\,
 2 \varepsilon_0^2 c^2
 E_{0}^2 \cos[k_0(z-z')-\omega_0 \tau] \qquad \\
 \times \left\langle \delta E_x({\bf r},t)\delta E_x({\bf r}',t+\tau)\right\rangle  ,
\label{intcorr02}
 \end{multline}
which depends on the two-point correlation function $\langle \delta E_x({\bf r},t)\, \delta E_x({\bf r}',t+\tau)\rangle$ of the fluctuating vacuum field at the surface of the detector.

We now introduce {Eq.~}(\ref{intcorr01}) and {Eq.~}(\ref{intcorr02}) into the expression for the PSD in {Eq.~}(\ref{sppwm2y}) which yields
\begin{multline}
S_{PP}(\omega) = \bar{P}^2\delta(\omega)+
2 \varepsilon_0 c\frac{\bar{P}}{A}\int_A da \int_{A}da'  \\
  \times[S_{EE}({\bf r}, {\bf r}',\omega\!+\!\omega_0) \,+\, S_{EE}({\bf r}, {\bf r}',\omega\!-\!\omega_0)],  \label{spppsd}
\end{multline}
where $\bar{P}= A I_0 = (A/2) \varepsilon_0 c E_0^2$ is the power of the incident field, $\delta(\omega)$ is the Dirac delta function, and $S_{EE}\equiv S_{E_x E_x}$, see Eq.(\ref{coh02de2}). To proceed we need to derive $S_{ E E}$ associated with vacuum fluctuations. 

We integrate the two-point correlation function (\ref{twopointcorr}) over the detector plane ($z=0$) and obtain 
\begin{multline}
\int_A da \int_A da' \,   \langle \delta{E}_x({\bf r},t)\, 
\delta {E}_x({\bf r}',t+\tau) \rangle\\
= \frac{\hbar c  A}{8 \pi \varepsilon_0}\int_0^{\infty} \!\!d k_z k_z\;  \big({\rm e}^{\imu  c k_z \tau} + {\rm c.c.}\big )
\end{multline}
where we restricted the integration domain to $k_z\in [0,\infty]$ to include the fields that \textit{propagate towards the detector}. This immediately leads to
\begin{equation}
    \int_A da \int_A da'  \;S_{EE}(\mathbf{r},\mathbf{r}',\omega)=\frac{\hbar |\omega| A}{8 \pi \varepsilon_0 c}. \label{seeres}
\end{equation}
Inserting Eq.~(\ref{seeres}) into Eq.~(\ref{spppsd}) yields
\begin{align}\notag
S_{PP}(\omega)  &= \bar{P}^2\, \delta(\omega)\\
&+2 \varepsilon_0 c\, \bar{P}  \left(\frac{\hbar|\omega-\omega_0|}{8 \pi c \varepsilon_0} 
+ \frac{\hbar|\omega+\omega_0|}{8 \pi c \varepsilon_0}\right ),
\end{align}
which, for frequencies  $|\omega|\ll\omega_0$, reduces to
\begin{eqnarray}
S_{PP}(\omega) \simeq \bar{P}^2\,\delta(\omega)+
\bar{P}\, \frac{\hbar\omega_0}{2 \pi} \, .
\label{spppsd3}
\end{eqnarray}
This result is consistent with W. Schottky's analysis based on the assumption of discrete detection events of energy $\hbar\omega_0$~\cite{schottky18}. As shown in Appendix A, the result (\ref{spppsd3}) can be equivalently derived using the fluctuation dissipation theorem (\ref{coh02de2}).

\section{Heterodyne Measurement of a Suspended Mirror \label{suspmirror}}
Based on the results of the previous section we now proceed to analyzing the system of interest  illustrated in Fig.~\ref{heterodyneillustration}. A deterministic electromagnetic field ${\bf E}_0$ with center frequency $\omega_0$ is reflected (scattered) from a suspended perfect mirror (reflection coefficient $r=1$). The scattered field ${\bf E}_{\rm sc}$ is combined with a frequency-shifted (deterministic) reference field ${\bf E}_{\rm ref}$ with center frequency $\omega_0+\Delta$. The fields are then directed on a photodetector that measures the power $P$, from which we derive an expression for the PSD. A stochastic background field $\delta {\bf E}$, originating from thermal and zero-point fluctuations, is superimposed to all fields. The mirror is characterized by the susceptibility $\chi(\omega)$, which transduces the radiation pressure force into a mirror displacement $z_0+\delta z_\text{rp}$, where $z_0$ originates from the response to the deterministic field and $\delta z_\text{rp}$ from the response to the stochastic electromagnetic field. The parameters of the mirror are its mass $m$, its mechanical eigenfrequency $\Omega_0$ and its damping $\gamma$.  Moreover, an additional stochastic displacement $\delta z_\text{th}$ accounts for fluctuations of the mirror originating from a coupling to an external thermal bath, e.g. residual gas at temperature $T$. In the following we choose the origin of the coordinate system to coincide with the static mirror displacement $z_0$ so that the mirror displacement  $\delta z=\delta z_\text{rp}+\delta z_\text{th}$ originates exclusively from fluctuations.

A mirror displacement of $\delta z$ changes the optical path length by $2\:\!\delta z$ and hence the reflected (scattered) field becomes 
\begin{eqnarray}
{\bf E}_{\rm sc}(z,t) &=& - E_0\, {\rm Re} \big[
{\rm e}^{-{\rm i} k_0
(z-2 \delta z)}\, {\rm e}^{-\imu \omega_0 t}\big ]\, {\bf n}_{x} \;  .
\end{eqnarray}
At the location of the detector the  scattered field (${\bf E}_{\rm sc}$) and the reference field (${\bf E}_{\rm ref}$) read as
\begin{align}
{\bf E}_{\rm sc}(t) &=- E_0\, \cos[2 k_0 \delta z - \omega_0 t] {\bf n}_{x},  \label{reffldif} \\[0.5ex]
{\bf E}_{\rm ref}(t) &= X \cos[(\omega_0+\Delta) t]{\bf n}_{x} +
Y\sin[(\omega_0+\Delta) t] {\bf n}_{x}  ,\label{reffldif2}
 \end{align}
where we suppressed common phase terms and utilized the quadratures $X$ and $Y$ in the expression for the reference field. For small mirror displacements ($k_0 \delta z\ll1$) we expand  to first order and obtain
\begin{equation}
{\bf E}_{\rm sc}(t) =- E_0\big[\cos(\omega_0 t) +
2 k_0 \:\!\delta z(t)\sin(\omega_0 t)\big] {\bf n}_{x} ,
\label{reffld}
 \end{equation}
where we have allowed the displacement $\delta z$ to be time dependent. This is valid in the adiabatic limit $\Omega_0 \ll \omega_0$, that is at any instant of time the mirror can be regarded as being fixed. According to {Eq.~}(\ref{reffld}) the position fluctuations are imprinted on the phase quadrature of the field.

We introduce the Fourier transform defined as ${\bf E}({\bf r},\omega)\equiv (2\pi)^{-1}\!
\int_\mathbb{R}d t\, {\bf E}({\bf r},t)\exp (\imu \omega t)$ so that the Fourier transform of the intensity (\ref{intdefmy}) can be represented as
\begin{equation}
I(x,y,\omega) \,=\,  \varepsilon_0 c \int_\mathbb{R} d\omega' \;
{{\bf E}}(x,y,\omega')\cdot{{\bf E}}(x,y,\omega-\omega'),
\label{ftint}
\end{equation}
which allows us to express the PSD (\ref{sppwm2y}) as
\begin{multline}
S_{PP}(\omega) \\
= \int_{A} da \int_{A} da'  \int_\mathbb{R}  d\omega'\, \big\langle I(x,y,\omega)\, I^{*}(x',y',\omega') \big\rangle  \label{radpf7+4}  .
\end{multline}
In order to obtain the statistical average experimentally, one records a time trace $P(t)$ over a finite time period and then evaluates the Fourier transform ${P}(\omega)$ and the product ${P}(\omega) \,{P}^{*}(\omega')$. This procedure is repeated many times, and at the end one takes the average $\langle  {P}(\omega)\, {P}^{*}(\omega')\rangle$ of the individual results. Note that in the following we are repeatedly using the fact that for a statistically stationary process different frequency components do not correlate, as dictated by the Wiener-Khinchin theorem. This can formally be expressed as $\langle {P}(\omega)\, {P}^{*}(\omega')\rangle \propto \delta(\omega\!-\!\omega')$.

\subsection{Evaluation of the detector signal}
 The electric field at the detector is composed of three parts
\begin{equation}
 {\bf E}({\bf r},t) =  {\bf E}_{\rm sc}({\bf r},t) \,+\, {\bf E}_{\rm ref}({\bf r},t)\,+\,\delta {\bf E}({\bf r},t)  ,
\label{totfldz}
\end{equation}
where ${\bf E}_{\rm sc}$ is the field scattered (reflected) from the mirror (\ref{reffld}),  ${\bf E}_{\rm ref}$ is the frequency-shifted reference field (\ref{reffldif2}), and $\delta {\bf E}$ are the background fluctuations. After inserting the fields into {Eq.~}(\ref{intdefmy}), accounting for the finite response time of the detector, and suppressing terms quadratic in fluctuations, we obtain an expression for the intensity $I\equiv \varepsilon_0 c (i_1/2+i_2+i_3$). The first term $i_1$ denotes the deterministic intensity of the reference beam and the reflected beam, respectively, with frequency components at $\omega=0$ and $\omega=\pm\Delta$, namely
\begin{multline}
    i_1(\omega)=(E_0^2+E_+E_-)\delta(\omega)\\
    -E_0E_-\delta(\omega+\Delta)-E_0E_+\delta(\omega+\Delta)\, ,
\end{multline}
where we have defined $E_\pm \equiv X \pm \imu Y$. The second term $i_2$ denotes the terms responsible for shot noise, that is
\begin{multline}
    i_2(\omega)=-E_0\, \delta E_x(\omega-\omega_0)-E_0\, \delta E_x(\omega+\omega_0)\\
    +E_+\, \delta E_x(\omega-\omega_0 -\Delta)+E_-\, \delta E_x(\omega+\omega_0 +\Delta).
\end{multline}
Finally, $i_3$ denotes the interferometric signal of the mirror's position, the actual measurement
\begin{align}
    i_3(\omega)=-\imu k_0 E_0 [E_-\, \delta z(\omega+\Delta)-E_+\,\delta z(\omega-\Delta)].
\end{align}
To ease the notation we dropped the $x,y$ dependence in the arguments of  $\delta E_x$. The PSD of the detector signal is calculated according to Eq.(\ref{radpf7+4}) and yields four terms
\begin{equation}
S_{PP}(\omega)  = S_{PP}^{(1)}(\omega) +S_{PP}^{(2)}(\omega) + S_{PP}^{(3)}(\omega) + S_{PP}^{(4)}(\omega).  
\label{fourterms}
\end{equation}
The first term is the deterministic signal, the second the shot noise, the third  the signal (mirror motion), and the fourth is a cross term between signal and shot noise (product of $i_2$ and $i_3$). It is the latter that gives rise to the sideband asymmetry. The remaining two cross terms (product of $i_1$ and $i_2$ and product of $i_1$ and $i_3$) are zero because they are linear in fluctuations  and average to zero. In the following we derive each of the four terms separately. 

\subsubsection{Deterministic signal:}
Using the powers of incident and reference beams $\bar{P}= (A/2) \varepsilon_0 c E_0^2$ and $\,\bar{P}_{\rm ref}  \equiv (A/2) \varepsilon_0 c |E_+|^2$, respectively, we obtain
\begin{multline}
    S_{PP}^{(1)}\simeq
    \bar{P}_{\rm ref}^2 \,\delta(\omega) + \bar{P} \bar{P}_{\rm ref}\, \delta(\omega+\Delta) + \bar{P} \bar{P}_{\rm ref} \,\delta(\omega-\Delta),
\end{multline}
where we made use of the fact that only same-frequency components correlate and assume now and in the following that the power of the reference beam is larger than the power of the incident beam ($\bar{P}_\text{ref} \gg \bar{P}$).

\subsubsection{Shot noise:}
The intensity correlation function of the shot noise terms $(i_2)$ reads
\begin{align}\notag
&\langle {i}_2(\omega)\, {i}_2^{*}(\omega') \rangle  = E_0^2 \, \langle \delta E_x(\omega-\omega_0) \, \delta E_x^{*}(\omega'-\omega_0)\rangle \\ \notag
&+ E_0^2\, \langle \delta E_x(\omega+\omega_0)\, \delta E_x^{*}(\omega'+\omega_0)\rangle\\ \notag
 &+ E_+E_- \, \langle \delta E_x(\omega-\omega_0-\Delta) \, \delta E_x^{*}(\omega'-\omega_0-\Delta)\rangle \nonumber  \\  \label{intcorrshn} 
 &+   E_+E_- \, \langle \delta E_x(\omega +\omega_0+\Delta) \,\delta E_x^{*}(\omega'+\omega_0+\Delta) \rangle\, .
\end{align}
The corresponding PSD becomes
 \begin{align}   \nonumber
S_{PP}^{(2)}(\omega)  &=
 \frac{\bar{P}}{4 \pi}  \hbar \, ( |\omega-\omega_0| +|\omega+\omega_0|)  \\
 &+ \frac{\bar{P}_{\rm ref}}{4 \pi} \hbar  \,( |\omega-\omega_0-\Delta| +|\omega+\omega_0+\Delta|)  ,
  \end{align}
where we made use of Eq.~(\ref{seeres}). For frequencies $|\omega \pm \omega_0| \simeq \omega_0$ and $|\omega \pm (\omega_0+\Delta)|\simeq \omega_0$ we find
\begin{eqnarray}
S_{PP}^{(2)}(\omega) \simeq \frac{\bar{P}_{\rm ref}}{2 \pi}\, \hbar\omega_0.
\label{shotnoisePSD} 
\end{eqnarray}

\subsubsection{Interferometric signal :}
The intensity correlation function generated by the interference of the reference field $\bf{E}_{\rm ref}$ and the scattered field $\bf{E}_{\rm sc}$ is
\begin{multline}
\langle {i}_3(\omega)\, {i}_3^{*\!}(\omega') \rangle  =  k_0^2 E_0^2 E_+E_- \\
\times \big[
 \langle {\delta z}(\omega\!+\!\Delta) \, {\delta z}^{*}(\omega'\!+\!\Delta) \rangle  \,+\,  \langle {\delta z}(\omega\!-\!\Delta) \, {\delta z}^{*}(\omega'\!-\!\Delta) \rangle
 \big]\, .
\end{multline}
The PSD becomes
\begin{align} \nonumber
S_{PP}^{(3)}(\omega)  &= 4 k_0^2 \bar{P} \bar{P}_{\rm ref}    \int_\mathbb{R} d\omega'  \big\langle {\delta z}(\omega-\Delta)\, {\delta z}^{*}(\omega'-\Delta)\big\rangle \\ \label{psdshnt2}
&+  4 k^2 \bar{P} \bar{P}_{\rm ref}   \int_\mathbb{R}  d\omega' \big\langle {\delta z}(\omega+\Delta)\, {\delta z}^{*}(\omega'+\Delta)\big\rangle.
\end{align}
The two integrals in {Eq.~}(\ref{psdshnt2}) correspond to the PSD of the mirror displacement which we write as
\begin{equation}
\delta z(t) = \delta z_\text{rp}(t)+\delta z_\text{th}(t) .
\label{zoind}
\end{equation}
In terms of the susceptibility (\ref{chi}) we can represent {Eq.~}(\ref{zoind}) in the frequency domain as
\begin{equation}
\delta {z}(\omega) = \chi(\omega)\, [{F}_\text{rp}(\omega)+{F}_\text{th}(\omega)],
\label{zzzz}
\end{equation}
where
$ {F}_\text{rp}$ is the Fourier transform of the radiation pressure force. It can be expressed as~\cite{novotny12}
\begin{equation}
{F}_\text{rp}(\omega)= \frac{2}{c} \int_{A_{\rm m}}\! da \: \delta I(x,y,\omega) ,
\label{radpf0tp}
\end{equation}
with $A_{\rm m}$ being the mirror area and $\delta I$ the Fourier transform of the fluctuating incoming intensity, namely
\begin{align}
    \delta I (\omega)&=\varepsilon_0 c \, E_0\, \delta E_x(\omega+\omega_0)+\varepsilon_0 c \,E_0\, \delta E_x(\omega-\omega_0).
    \label{radprforcev}
\end{align}
Again using {Eq.~}(\ref{seeres}) it immediately follows that the PSD of the force reads
\begin{eqnarray}
S^\text{rp}_{ F F}   =\frac{4\hbar\omega_0}{2 \pi c^2} \bar{P},
 \label{radpf9+++}
\end{eqnarray}
which is the radiation pressure shot noise. 

$S^\text{rp}_{ F  F}$ heats the motion of the mirror while the intrinsic damping $\gamma$ in Eq.~(\ref{chi}) cools it. The steady state energy  $E_{\infty} = m\Omega_0^2  \langle \delta z_{\rm rp}^2 \rangle$ is calculated as
\begin{multline}
E_{\infty} = m\Omega_0^2 \int_{-\infty}^{\infty} \!\!\! d\omega\, \left|\chi(\omega)\right|^2  S^\text{rp}_{ F  F} =\frac{2\:\!\bar{P}}{m c^2}  \frac{\hbar\omega_0}{\gamma}\; ,\;\;\;\label{theory12}
\end{multline}
where we have used {Eq.~}(\ref{chi}) and 
{Eq.~}(\ref{radpf9+++}). Introducing the photon recoil heating rate $\Gamma$ defined as the energy added to the mirror (in units of $\hbar\Omega_0$) per unit time we can express the steady-state energy as the ratio of heating and cooling as $E_{\infty}= \hbar\Omega_0\, \Gamma/\gamma$. With the help of 
{Eq.~}(\ref{theory12}) we then obtain
\begin{equation}
\Gamma = \frac{4\:\!\bar{P}}{2 m c^2} \frac{\omega_0}{\Omega_0} \; ,
\label{theory14}
  \end{equation}
which is also referred to as the quantum backaction rate.

Since $ F_\text{rp}$ and $\delta z_\text{th}$ are uncorrelated, Eqs.~(\ref{zzzz}) and (\ref{radpf9+++}) yield 
\begin{equation}
S_{zz}(\omega)  = |\chi(\omega)|^2 (S^\text{rp}_{  F F}  + S^\text{th}_{ F F}) ,
\label{psdzz}
\end{equation}
where $S^\text{th}_{ F F}$ is given by the FDT in {Eq.~}(\ref{fluctuations}). Equation (\ref{psdshnt2}) now yields
\begin{align}\notag
 S_{PP}^{(3)}(\omega) &= 4k_0^2\bar{P}\bar{P}_\text{ref}\big[\:\!|\chi(\omega+\Delta)|^2+|\chi(\omega-\Delta)|^2\big]\\
&\times(S^\text{rp}_{ F  F}+S^\text{th}_{ F  F}).
\end{align}
The expression consists of two noise sidebands at $\omega=\Delta\pm\Omega_0$ of equal amplitude. Both of these sidebands are made of two contributions, one stemming from the intrinsic position fluctuations of the mirror ($\delta z_\text{th}$) and another induced by  radiation pressure shot noise ($\delta z_\text{rp}$). The latter is referred to as measurement backaction. For a high-Q oscillator $(\gamma\ll\Omega_0)$,  and a frequency shift much larger than the oscillation frequency ($\Delta\gg\Omega_0$), {we arrive at}
\begin{eqnarray}\label{eq:S3}
S_{PP}^{(3)}(\Delta\pm\Omega_0)\simeq 4k_0^2 \bar{P}\bar{P}_{\rm ref}  \frac{2z^2_{\rm zp}}{\pi \gamma}\left( {\frac{\Gamma}{\gamma}+\bar{n}+\frac{1}{2}}\right), 
\end{eqnarray}
where $z^2_{\rm zp}= \hbar/(2 m \Omega_0)$ is the mean-square amplitude of the zero-point motion,
 and $\bar{n}\equiv[\exp(\hbar\Omega_0/k_B T)-1]^{-1}$ is the mean thermal occupation number. Finally, for  $k_B T /\ \hbar\Omega_0\ll~1$ we arrive at
\begin{eqnarray}
	S_{PP}^{(3)}(\Delta\pm\Omega_0)\simeq 4k_0^2 \bar{P}\bar{P}_{\rm ref}  \frac{2z^2_{\rm zp}}{\pi \gamma}\left(\frac{\Gamma}{\gamma}+ \frac{1}{2}\right).
	\label{asymtw}
\end{eqnarray}

\subsubsection{Correlation between imprecision and backaction :}
The last term contributing to the intensity correlation function is a cross term between $i_2$ and $i_3$, that is,
\begin{multline}
    \langle {i}_2(\omega) \,{i}_3^{*}(\omega') \rangle+\langle {i}_3(\omega)\, {i}_2^{*}(\omega') \rangle=-2 k_0 E_0 E_+ E_-\\
    \times {\rm Im} [    \big\langle\delta E_x(\omega + \omega_0 +\Delta)\,  {\delta z}^{\ast\!}(\omega' +\Delta)\rangle \\
    -\langle\delta E_x(\omega-\omega_0-\Delta)\,  {\delta z}^{\ast\!}(\omega'-\Delta) \big\rangle].\label{crosscorr3} 
\end{multline}

In order to evaluate {Eq.~}(\ref{crosscorr3}), let us first derive an expression for the correlation function $\langle \delta E_x(\omega\pm (\omega_0\pm\Delta))\delta z^*(\omega'\pm\Delta)\rangle$. Introducing {Eq.~}(\ref{zzzz}), and using the fact that $\delta E_x$ and $\delta{z}_\text{th}$ are uncorrelated $\langle \delta E_x\delta{z}^*_\text{th}\rangle =0$ leads to
\begin{multline}\label{eq:intermediateStep1}
\langle\delta E_x(\omega\pm\omega_0)\delta z^*(\omega')\rangle=\chi^*(\omega')\langle\delta E_x(\omega\pm\omega_0)F^*_\text{rp}(\omega')\rangle\\
=2\epsilon_0E_0\chi^*(\omega')\int_{A_m}da\,\langle\delta E_x(\omega\pm\omega_0)\delta E^*_x(\omega'\pm\omega_0)\rangle,
\end{multline}
where we have shifted both arguments by $\pm\Delta$ for better readability. The integral  in Eq.~(\ref{eq:intermediateStep1}) can directly be evaluated using {Eq.~}(\ref{seeres}), which yields
\begin{multline}\label{eq:intermediateStep2}
    \langle\delta E_x(\omega\pm(\omega_0+\Delta))\delta z^*(\omega'\pm\Delta)\rangle=\\
  \frac{\hbar |\omega\pm (\omega_0+\Delta)| E_0}{4\pi c}\chi^*(\omega\pm\Delta)\delta(\omega-\omega').
\end{multline}
Inserting {Eq.~}(\ref{eq:intermediateStep2}) in {Eq.~}(\ref{crosscorr3}) immediately leads to the final expression 
\begin{equation}
S_{PP}^{(4)}(\omega)  = 4k_0^2 \bar{P}\bar{P}_{\rm ref}\, \frac{\hbar}{2\pi} {\rm Im} \big[
\chi(\omega +\Delta)  -\chi(\omega-\Delta) \big],
\end{equation}
for frequencies $|\omega \pm (\omega_0+\Delta)|\simeq \omega_0$. Again considering a high-Q oscillator $(\gamma\ll\Omega_0)$ and a frequency shift much larger than the oscillation frequency ($\Delta\gg\Omega_0$) we obtain
\begin{eqnarray}
S_{PP}^{(4)}(\Delta\pm\Omega_0) \,\simeq\, \mp\, 4k_0^2 \bar{P}\bar{P}_{\rm ref}  \frac{z^2_{\rm zp}}{\pi \gamma} , 
\end{eqnarray}
that is, one sideband has a positive amplitude whereas the other one has a negative amplitude. When combined with $S_{PP}^{(3)}$ we find that one sideband gets reduced and the other one  increased in amplitude.

The derivation of the motional sideband asymmetry using stochastic electrodynamics is the main result of this article. In the following we show how this result manifests itself in balanced photodetection.

\subsection{Balanced Photodetection}
\begin{figure}[!t]
\begin{center}
\vspace{1.0ex}
\includegraphics[width=0.48\textwidth]{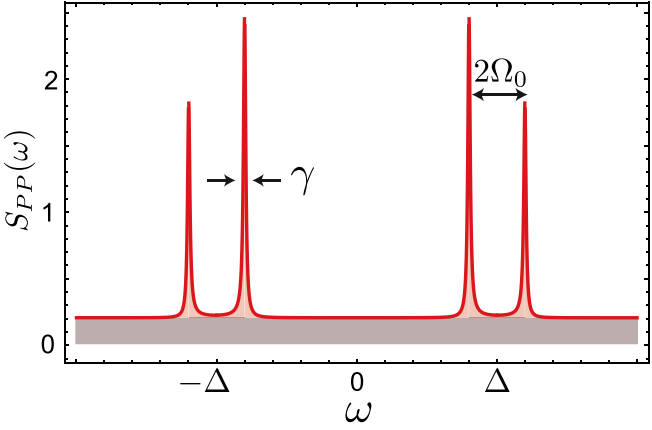}
\quad\vspace{-2em}
\end{center}
\caption{Heterodyne spectrum $S_{PP}(\omega)$ for $k_B T= 3\hbar\Omega_0$ featuring a sideband asymmetry. The signal (red) is on top of a background due to shot noise (grey). It is assumed that $\gamma \ll \Omega_0 \ll \Delta$. \\}
\label{hetspecfig}
\end{figure}

Balanced detection consists of measuring a differential signal with a pair of photodetectors. This eliminates the signal that is common to both photodetectors, that is, any DC signal or classical variations in laser power (relative intensity noise). Combining the four contributions of the PSD and rejecting all deterministic terms yields
\begin{align}\notag 
    &S_{PP}(\omega)\,=\,\frac{\bar{P}_\text{ref}}{2\pi}\,\hbar \omega_0\\\notag
    &+4k_0^2 \bar{P}\:\!\bar{P}_\text{ref}\,[\:\! |\chi(\omega+\Delta)|^2 +|\chi(\omega-\Delta)|^2 ](S^\text{rp}_{ F F}+S^\text{th}_{ F  F})\\ \label{sidebandfinal}
    &+4k_0^2 \bar{P}\:\!\bar{P}_\text{ref}\,\frac{\hbar}{2\pi}\, \text{Im}[\chi(\omega+\Delta)-\chi(\omega-\Delta)]\, .
\end{align}
We recall that $\bar{P}$ and $\omega_0$ are the power and frequency of the laser incident on the mirror, and  $\bar{P}_\text{ref}$ and $\omega_0+\Delta$ are the power and the frequency of the reference beam, respectively. The result (\ref{sidebandfinal})  is expressed in terms of the  susceptibility $\chi(\omega)$ in Eq.~(\ref{chi}), the radiation pressure shot noise $S^\text{rp}_{ F F}$ in Eq.(\ref{radpf9+++}) and the PSD of the mirror fluctuations $S^\text{th}_{ z z}=|\chi|^2 S^\text{th}_{ F  F}$ in Eq.~(\ref{coh02de3q}). The first line is the shot noise (assuming $\Delta\ll\omega_0$ and $\bar{P}\ll\bar{P}_\text{ref}$) and the second and third lines are the heterodyne sidebands, which differ in magnitude due to  correlations between measurement and backaction.  The difference in amplitudes is proportional to twice the PSD of the mirror's zero-point motion 
$\lim_{T\to0}S^\text{th}_{ z z}(\Omega_0) = z^2_{\rm zp}/ (\pi \gamma)= \hbar/(2 \pi  m \gamma \Omega_0)$, a phenomenon referred to as sideband asymmetry. As an example, we show in Fig.~\ref{hetspecfig} the PSD for the case of $k_B T= 3\hbar\Omega_0$.

\section{Discussion \label{discussionpsd}}
In the analysis presented here, the sideband asymmetry has its origin in the cross-term $S_{PP}^{(4)}(\omega)$ and can be traced back to the correlation between two pathways for field fluctuations to reach the detector, a direct path (represented by $i_2$) and an indirect path via reflection from the mirror (represented by $i_3$) . Field fluctuations that reach the detector via the indirect path impart radiation pressure on the mirror and the resulting displacement gives rise to optical path length modulation. For the zero-temperature limit this path length modulation corresponds to the zero-point motion of the mirror, resulting in a motional sideband asymmetry of  $2\times \lim_{T\to0}S^\text{th}_{ z z}(\Omega_0)$. This viewpoint agrees with the interpretation by Weinstein et. al.~\cite{weinstein14}. 

Note that the sideband asymmetry is independent of temperature, that is, the difference between the two sidebands amounts to $2 \times z^2_{\rm zp}/ (\pi \gamma)$ no matter how hot the mirror is. This difference can be used as a ruler to determine the temperature of the mirror. To see this we note that, for sufficiently small shot noise, the amplitude of the blue sideband at $\omega=\Delta +\Omega_0$ (anti-Stokes sideband) is proportional to the mean thermal occupation number $\bar{n}$ (c.f. Eq.~(\ref{eq:S3})) , whereas the amplitude of the red sideband at $\omega=\Delta -\Omega_0$ (Stokes sideband) is, proportional to $\bar{n}+1$ with the same proportionality constant. Thus, by means of the sideband difference we can determine $\bar{n}$ and hence the temperature of the mirror. Determining the temperature of a thermal bath by means of the sideband asymmetry is denoted {\em sideband thermometry}~\cite{purdy15}.

Let us now introduce the total PSD $S^{\rm tot}_{zz}$ describing the mirror displacement $\delta z$. It defines the mean-square displacement as our result for $S_{PP}$ defines the minimum mirror displacement that can be measured, that is,
\begin{equation}
\langle \delta z^2\rangle_{\rm  }=  \int_\mathbb{R} d\omega \, S^{\rm tot}_{zz}(\omega)  =  (4k_0^2 \bar{P} \bar{P}_{\rm ref})^{-1} \!\int_\mathbb{R} d\omega \, S_{PP}(\omega) ,
\label{stotzz}
\end{equation}
and can be calculated using {Eq.~}(\ref{sidebandfinal}). In the derivation of  {Eq.~}(\ref{sidebandfinal}) we assumed that the power $\bar{P}$ that is received by the photodetector is the same as the power that is incident on the mirror (the $\bar{P}$ in the expression of $S_{ F F}^\text{rp}$). However, due to imperfect detection (photon losses, photon absorption, finite detector efficiency), this might not be the case. To account for such imperfect detection we introduce the detection efficiency $\eta\in [0,1]$ and substitute all the $\bar{P}$ in  {Eq.~}(\ref{sidebandfinal}) by $\eta \bar{P}$ while leaving the $\bar{P}$ in the expression of $S_{ F F}^\text{rp}$ unchanged. With the help of {Eq.~}(\ref{sidebandfinal}), {Eq.~}(\ref{stotzz}) and defining the measurement noise  ${S}^\text{im}_{zz}(\omega)\equiv \hbar \omega_0 /(8 \pi k_0^2 \eta \bar{P})$, we then obtain
\begin{align}\notag 
    S_{zz}^\text{tot}(\omega)&=S_{zz}^\text{im}(\omega)+S^{\rm fluct}_{zz}(\omega)\\
    &+[|\chi(\omega+\Delta)|^2 +|\chi(\omega-\Delta)|^2 ]\, S^\text{rp}_{ F  F},
\end{align}
where we defined
\begin{align}
S^{\rm fluct}_{zz}(\omega) &\equiv [|\chi(\omega+\Delta)|^2 + |\chi(\omega-\Delta)|^2]{S}^\text{th}_{ F F}\\
&+\frac{\hbar}{2\pi} {\rm Im} [\chi(\omega+\Delta)-\chi(\omega-\Delta)].
\end{align}
This term describes intrinsic fluctuations (thermal and zero-point) that are independent of laser power $\bar P$. On the other hand,  measurement noise ${S}^\text{im}_{zz}$ scales inversely with $\bar P$  and radiation pressure shot noise $S^\text{rp}_{ F F}$ scales linearly with $\bar{P}$ , see Eq.(\ref{radpf9+++}). Therefore, there is an optimum power $\bar{P}$ that minimizes the total noise $S^{\rm tot}_{zz}$. This minimum occurs when measurement noise equals radiation pressure shot noise, that is,
\begin{eqnarray}
{S}^\text{im}_{zz}(\omega)  = [ |\chi(\omega+\Delta)|^2 + |\chi(\omega-\Delta)|^2]\,S^\text{rp}_{ F F} (\omega),
\end{eqnarray}
a condition referred to as standard quantum limit (SQL). Because $\gamma\ll\Delta$ the spectra of $\chi(\omega \pm \Delta)$ do not overlap and hence one can only be resonant with one of the terms in the bracket. 
Using ${S}^\text{im}_{zz} = \hbar \omega_0 /(8 \pi k_0^2 \eta \bar{P})$ and 
$S^\text{rp}_{ F F} = 4\hbar\omega_0  \bar{P} /(2 \pi c^2)$ we find 
\begin{eqnarray}
{S}^\text{im}_{zz} \, S^\text{rp}_{ F  F} = \frac{1}{\eta}\frac{\hbar^2}{4 \pi^2}.
\label{heisenbergheterod}
\end{eqnarray}
The inclusion of thermal noise makes this product worse and in general ${S}^\text{im}_{zz} S^\text{rp}_{ F  F} \ge \hbar^2/(4\pi^2\eta)$. This Heisenberg relation is a factor of four worse than the Heisenberg relation for homodyne detection ($\Delta = 0$) (c.f. Appendix B). The difference is analogous to the well known signal-to-noise difference between homo- and heterodyne detection in spectroscopy. 

In Appendix C we use the input-output formalism to show that a quantum framework leads to the same result as Eq.~(\ref{sidebandfinal}). In general, we can expect stochastic electrodynamics to yield the same predictions of quantum mechanics as long as the dynamics of the system and the measurement process are linear. Indeed, in order to demonstrate a genuine quantum phenomenon---a phenomenon that cannot be explained by classical mechanics even after the introduction of zero point fluctuations---we must introduce nonlinearities in the system, as discussed in Appendix C.

\section{Conclusions \label{conclusionspsd}}
Using stochastic electrodynamics and a linearized theory of light-matter interaction we have derived the heterodyne spectrum resulting from the reflection of a laser beam from a suspended mirror. We find that the sideband asymmetry results from the correlation of two ways that field fluctuations can reach the detector: a direct path and an indirect path via reflection
from the mirror. Our result for the power spectral density of the detected power agrees with the results obtained by other methods, such as input-output theory based on quantum Langevin equations~\cite{weinstein14} or quantum linear response theory~\cite{khalili12} (c.f. Appendix C). We stress that our calculation requires no quantum mechanics and that a sideband asymmetry can be quantitatively explained by classical means if one assumes zero-point fluctuations, that is that light and matter fluctuate at zero temperature. The theory outlined here can be readily extended to the motion of other systems that respond linearly to the electromagnetic field (e.g. levitated particles~\cite{millen20,gonzales21}) and adjusted for finite detection efficiencies and finite gas pressures. The Raman sideband problem studied in this paper adds to the list of problems that can be successfully treated by stochastic electrodynamics.

\begin{acknowledgments}
L.N. thanks Carsten Henkel, Girish Agarwal and Felix Tebbenjohanns for valuable discussions and input. This work has been supported by ETH Z{\"u}rich. 
\end{acknowledgments}

\clearpage

\section*{Appendix A: Detector Shot Noise derived via FDT}
We make use of the FDT (\ref{coh02de2}) and the {\em equilibrium} scalar Green function
\begin{eqnarray}
G_0({\bf r},{\bf r}',\omega) &=& \frac{{\rm e}^{\imu k |{\bf r}-{\bf r}'|}+{\rm e}^{-\imu k |{\bf r}-{\bf r}'|}}{8\pi|{\bf r}-{\bf r}'|}.
\end{eqnarray}
The first term corresponds to the outgoing  Green function $G_0^{+}$ and the second term to the incoming  Green function $G_0^{-}$. The equal superposition ensures  that there is no net energy transport (equilibrium), i.e.~the time-averaged Poynting vector is zero at any point in space. The dyadic Green function is derived as ${\displaystyle \G({\bf r},{\bf r}',\omega)} = \left[ {\displaystyle {I}} + k^{-2} \nabla \otimes \nabla\right] G_0({\bf r},{\bf r}',\omega)$, where ${\displaystyle {I}}$ denotes the $3\times 3$ identity matrix and $\otimes$ the dyadic product. We consider only vacuum fields  \textit{propagating towards the detector} and hence the incoming part of the equilibrium Green function  $\G^{-}$ needs to be set to zero.

Using the angular spectrum representation of the Green function~\cite{novotny12} it immediately follows that
\begin{eqnarray}
\int_A da \, {\displaystyle G_{xx}^{+}({\bf r},{\bf r}',\omega)}
=\frac{\imu c}{4\omega}  .
\label{ch3105b} 
\end{eqnarray}
With the help of {Eq.~}(\ref{coh02de2}) we now obtain
\begin{eqnarray}
\int_A da\,S_{ E E}({\bf r}, {\bf r}',\omega) =  
 \frac{\mu_0 c }{4\pi } \left[
\frac{\hbar\omega}{2} \coth\left(\frac{\hbar\omega}{2 k_B T} \right)
\right].
\label{eq22}
\end{eqnarray}
Inserting Eq.~(\ref{eq22}) into Eq.~(\ref{spppsd}), assuming that $|\omega\pm \omega_0| \simeq \omega_0 $, and taking the zero-temperature limit yields
\begin{eqnarray}
S_{PP}(\omega) \simeq \bar{P}^2\,\delta(\omega)+
\bar{P}\, \frac{\hbar\omega_0}{2 \pi},
\label{spppsd4}
\end{eqnarray}
in agreement with the result (\ref{spppsd3}).\\

\section*{Appendix B: Homodyne Detection}
To obtain the solutions for the homodyne case we cannot simply set $\Delta=0$ in the final results since we assumed $\Delta\gg\Omega_0$. We therefore return to the intensity at the detector and set $\Delta=0$. This yields $I\equiv\varepsilon_0 c\, (i_1/2+i_2+i_3)$, where
\begin{align}
    i_1(\omega)&=(\tilde{X}^2+Y^2)\, \delta(\omega),\\
    i_2(\omega)&=(\tilde{X}+\imu Y)\, \delta E_x(\omega-\omega_0)+(\tilde{X}-\imu Y)\, \delta E_x(\omega+\omega_0),\\
    i_3(\omega)&=- 2k_0 E_0  Y \delta z(\omega),
\end{align} 
where we have defined $\tilde{X}\equiv X-E_0$. As before, the three terms ${i}_1$, $i_2$ and ${i}_3$ give rise to a PSD of a balanced homodyne detection scheme with four terms
\begin{align}\notag
    S^\text{hom}_{PP}(\omega)&\,=\,\frac{\tilde{P}_\text{ref}}{2\pi}\,\hbar\omega_0+16k_0^2 \bar{P}\tilde{P}_\text{ref}\sin(2\theta)\frac{\hbar}{4\pi}\text{Re}[\chi(\omega)]\\
    &+16k_0^2 \bar{P}\tilde{P}_\text{ref}\,\sin^{2\!}\theta\, |\chi(\omega)|^2 (S^\text{rp}_{ F F}+S^\text{th}_{ F F})\label{totpsd3},
\end{align}
where $\tilde{P}_\text{ref}\equiv (A/2)\varepsilon_0 c (\tilde{X}^2+Y^2)/2$, $\cos \theta \equiv \tilde{X}/(\tilde{X}^2+Y^2)^{1/2}$, and $\sin \theta \equiv -Y/(\tilde{X}^2+Y^2)^{1/2}$. Note that $\theta$ is the phase between the signal (field reflected from a suspended mirror) and the reference beam. Following the same steps as in the heterodyne detection scheme it immediately follows that for $\theta=\pi/2$ the product of imprecision and backaction is 
\begin{eqnarray}
{S}^\text{im}_{zz} \, S^\text{rp}_{ F  F} \,=\, \frac{1}{\eta} \frac{\hbar^2}{16 \pi^2} ,
\end{eqnarray}
which is a factor of four better than the Heisenberg relation for heterodyne detection (\ref{heisenbergheterod}). Note that angles $\theta\neq\pi/2$ can reduce $S^\text{hom}_{PP}$ below the shot-noise limit and give rise to ponderomotive squeezing. \\

\section*{Appendix C: Quantum Treatment of the Sideband Asymmetry}

In this section of the appendix we show that the results derived in the main article agree with a quantum-theoretical treatment of the problem. First, we briefly define and motivate the expression for the Hamiltonian that generates the dynamics between the suspended mirror and the electromagnetic field. Second, we derive the Langevin equation for the center-of-mass operator and the input-output relation for the quadratures of the electromagnetic field. Finally, we combine all results to derive an expression for the power spectral density (PSD) obtained by an ideal balanced heterodyne  (homodyne) detection scheme and show that it agrees with Eq.(\ref{sidebandfinal}) (Eq.(\ref{totpsd3})) in the article.

We consider a suspended mirror of mass $m$ and natural frequency $\Omega_0$ in the presence of a state of the electromagnetic where all modes are in vacuum except for a highly populated coherent mode at frequency $\omega_0$. In the system described by Fig.~\ref{heterodyneillustration}, the populated mode is a plane wave of angular frequency $\omega_0$ propagating along the $z$ axis. 
The total Hamiltonian $\hat{H}=\hat{H}_\text{cm}+\hat{H}_\text{em}+\hat{H}_\text{int}$ consist of three parts. The first term
\begin{equation}
    \hat{H}_\text{cm}=\hbar \Omega_0 \hat{b}^\dagger \hat{b},
\end{equation}
generates the free dynamics of the suspended mirror with operators that fulfill the bosonic commutation rules $[\hat{b},\hat{b}^\dagger]=1$. The center-of-mass operator reads $\hat{z}=z_\text{zp}(\hat{b}+\hat{b}^\dagger)$ with $z_\text{zp}= \sqrt{\hbar/(2 m \Omega_0)}$. The origin of the coordinate system is chosen to coincide with the mirror's equilibrium position in presence of the coherent beam. The second term 
\begin{equation}
    \hat{H}_\text{em}=\hbar\sum_{\epsilon} \int_{\mathbb{R}^3}\text{d}\mathbf{k}  \Delta(k) \hat{a}_\epsilon^\dagger(\mathbf{k})\hat{a}_\epsilon(\mathbf{k}),
\end{equation}
generates the free dynamics of the electromagnetic field in the presence of the mirror with operators that fulfill the bosonic commutation rules $[\hat{a}_\epsilon(\mathbf{k}),\hat{a}_{\epsilon'}(\mathbf{k}')]=\delta_{\epsilon\epsilon'}\delta(\mathbf{k}-\mathbf{k}')$. The indices $\epsilon\in \lbrace 1,2 \rbrace$ and $\mathbf{k}\in \mathbb{R}^3$ fully characterize each eigenmode of the electromagnetic field at frequency $\omega(k)=c \lvert \mathbf{k} \rvert$. Note that the Hamiltonian is defined in a reference frame rotating at frequency $\omega_0$ of the strongly populated coherent mode, where $\Delta(k)\equiv \omega(k)-\omega_0$. The third term generates the coupled dynamics induced by radiation pressure. For $(\langle \hat{z}^2\rangle)^{1/2}$ smaller than any relevant length scale associated to the electromagnetic fields, and in a frame that is displaced with respect to the coherent mode, the Hamiltonian can be linearized in $\hat{z}$ and reads
\begin{equation}
    \hat{H}_\text{int}=\hbar \sum_{\epsilon}\int_{\mathbb{R}^3}\text{d}\mathbf{k} [G_\epsilon(\mathbf{k})\hat{a}^\dagger_\epsilon(\mathbf{k}) +G_\epsilon^*(\mathbf{k}) \hat{a}_\epsilon(\mathbf{k})](\hat{b}+\hat{b}^\dagger).
\end{equation}
Note that a closed expression for the coupling strengths $G_\epsilon(\mathbf{k})$ can be derived and expressed in terms of the eigenmodes of the electromagnetic field~\cite{pirandola03,maurer22}.

It is now straightforward to derive Langevin equation for $\hat{z}$ by first deriving the Heisenberg equation $d\hat{z}/dt= [\hat{z},\hat{H}]/(\imu \hbar)$ in a time local form and including a phenomenological damping rate $\gamma$ due to coupling to an additional thermal environment (residual gas) at a temperature $T$ \cite{gardiner2004}. In the spectral domain we have
\begin{equation}\label{eq:spectrumposition}
    \hat{z}(\omega)=\chi(\omega)[\hat{F}_\text{rp}(\omega)+\hat{F}_\text{th}(\omega)],
\end{equation}
where the susceptibility reads $\chi(\omega)=[m(\Omega_0^2-\omega^2-\imu \gamma\omega)]^{-1}$. The above operator-valued equation corresponds to that of a driven harmonic oscillator with zero-mean driving $\langle \hat{F}_\text{rp}(\omega) \rangle = \langle \hat{F}_\text{th}(\omega)\rangle=0$. The first driving term $\hat{F}_\text{rp}(\omega)=\hbar\sqrt{\Gamma}[\hat{a}_\text{in}(\omega)+\hat{a}^\dagger_\text{in}(\omega)]/z_\text{zp}$ originates from the radiation-pressure induced interaction with a single collective mode of the electromagnetic field, the so called input \textit{interacting mode} \cite{militaru2022}
\begin{multline}
    \hat{a}_\text{in}(\omega)\equiv\lim_{t_0\to -\infty }\frac{1}{\sqrt{\Gamma}}\sum_{\epsilon} \int_{\mathbb{R}^3}\text{d}\mathbf{k} G^*_\epsilon(\mathbf{k})\\
    \times \hat{a}_\epsilon(\mathbf{k})e^{\imu \Delta(k) t_0}\delta(\omega+\Delta(k)),
\end{multline}
fulfilling the bosonic commutation rule $[\hat{a}_\text{in}(\omega),\hat{a}_\text{in}^\dagger(\omega')]=\delta(\omega+\omega')/(2\pi)$ and where $\Gamma$ denotes the photon recoil heating rate. The second driving term $\hat{F}_\text{th}(\omega)$ originates from the coupling to the additional thermal environment at temperature $T$. Note that the symmetrized correlation functions of the forces read
\begin{align}\label{eq:corr1}
    \frac{\langle \lbrace \hat{F}_\text{rp}(\omega) , \hat{F}_\text{rp}(\omega' )\rbrace \rangle}{2}&= \frac{\hbar^2 \Gamma}{2 \pi z_\text{zp}^2}\delta(\omega+\omega'),\\ \label{eq:corr2}
    \frac{\langle \lbrace \hat{F}_\text{th}(\omega) , \hat{F}_\text{th}(\omega' )\rbrace \rangle}{2}&= \frac{\hbar^2 \gamma}{2 \pi z_\text{zp}^2}\frac{\coth(\hbar \Omega_0/2k_b T)}{2}\delta(\omega+\omega'),
\end{align}
with $\lbrace \hat{A},\hat{B} \rbrace \equiv \hat{A}\hat{B}+\hat{B}\hat{A}$. Comparing Eq.(\ref{eq:corr1}) with Eq.(\ref{radpf9+++}) we can express the recoil heating rate in terms of the power $\bar{P}$ of the incoming coherent mode, that is $\Gamma = 2 \bar{P}/(m c^2) \times \omega_0/\Omega_0$ (c.f. Eq.~\ref{theory14}).

Before proceeding, let us point out differences and similarities between the classical and  quantum treatments of the case at hand. In both cases, the (quantum) Langevin equations lead to Eq.~\eqref{eq:spectrumposition} for the position in Fourier domain. The difference between the two frameworks lies in the fact that in the quantum case, $\hat{z}(t)$ represents an operator in the Heisenberg picture. As such, it does not in general commute with itself at different times, and any measurable quantity that is sensitive to this commutator differs from its classical counterpart. For linear dynamics, however, and for a detection system that measures only the symmetrized correlation function, any commutator vanishes from the measured signals. As long as Eqs.~\eqref{eq:corr1} and~\eqref{eq:corr2} hold, fluctuation electrodynamics delivers the same result that can be derived with the full quantum treatment of this section. In order to observe a measurable difference between the two formalism, we need to introduce nonlinearities in the equations of motion or in the measurement process. In this case, the commutator between operators does in general imprint a signature on the recorded trajectories.

The center-of-mass operator is imprinted in the quadratures of the electromagnetic field, as reflected in the input-output relation of the generalized quadrature $\hat{X}(\theta,\omega)\equiv \hat{a}(\omega)\exp(- i \theta)+\hat{a}^\dagger(\omega) \exp(i \theta)$~\cite{gardiner1985, militaru2022, magrini2022}, namely
\begin{equation}\label{eq:inout}
    \hat{X}_\text{out}(\theta,\omega)=\hat{X}_\text{in}(\theta,\omega)-\sqrt{4\Gamma}\sin(\theta)\frac{\hat{z}(\omega)}{z_\text{zp}},
\end{equation}
where the output interacting mode reads
\begin{multline}
    \hat{a}_\text{out}(\omega)\equiv\lim_{t_0\to \infty }\frac{1}{\sqrt{\Gamma}}\sum_{\epsilon} \int_{\mathbb{R}^3}\text{d}\mathbf{k} G^*_\epsilon(\mathbf{k})\\
    \times \hat{a}_\epsilon(\mathbf{k})e^{\imu \Delta(k) t_0}\delta(\omega+\Delta(k)),
\end{multline}
Let us now consider an \textit{ideal} balanced heterodyne detection scheme with a detector that measures the symmetrized correlation function, and a reference field with a large power $\bar{P}_\text{ref}$ and detuning $\Delta$ with respect to $\omega_0$. Choosing an ideal reference field gives access to the the following PSD of the generalized output quadrature \cite{tebbenjohanns2019}
\begin{multline}
    S_\text{het}^\text{tot}(\omega)\equiv  \frac{1}{2\pi} \int_\mathbb{R} d \tau \\
    \times \frac{\langle\langle\lbrace \hat{X}_\text{out}(\theta(t),t), \hat{X}_\text{out}(\theta(t+\tau),t+\tau\rbrace\rangle \rangle_t}{2}\exp(\imu \omega \tau).
\end{multline}
with a time-dependent phase $\theta(t)\equiv t \Delta  $. Note that the integrand is averaged in $t$ over a period $2\pi/\Delta$ to account for the finite response time $\tau \gg 2\pi/\Delta$ of the photodetector. It directly follows that
\begin{align}\notag
	S_\text{het}^\text{tot}(\omega) &= \frac{1}{4}\sum_{\sigma = \pm }S^\text{out}_{X_\theta X_\theta}(\omega+\sigma\Delta)+S^\text{out}_{Y_\theta Y_\theta}(\omega+\sigma\Delta)\\ 
    \notag
	&+\frac{\imu}{4} \sum_{\sigma = \pm }  \sigma [S^\text{out}_{X_\theta Y_\theta}(\omega+\sigma\Delta)-S^\text{out}_{Y_\theta X_\theta}(\omega+\sigma\Delta)],
\end{align}
where we have defined
\begin{equation}
    S^\text{out}_{X_\theta X_{\theta'}}(\omega)\equiv \int_{\mathbb{R}}d\omega' \frac{\langle \lbrace \hat{X}_\text{out}(\theta,\omega), \hat{X}_\text{out}(\theta',\omega') \rbrace \rangle}{2},
\end{equation}
with $\hat{Y}(\theta,\omega)=\hat{X}(\theta+\pi/2,\omega)$. Also note that an ideal balanced homodyne detection scheme gives access to the PSD of the generalized output quadrature $S_\text{hom}^\text{tot}(\omega)\equiv  S^\text{out}_{X_\theta X_{\theta}}(\omega).$ To derive an expression for both the heterodyne and homodyne case we use the input-output relation Eq.(\ref{eq:inout}) to arrive at
\begin{multline} \label{eq:inoutPSD}
    S^\text{out}_{X_\theta X_{\theta'}}(\omega)=\\
    S^\text{in}_{X_\theta X_{\theta'}}(\omega)+S_{zz}(\omega)+S^\text{in}_{X_{\theta} z}(\omega)+S^\text{in}_{z X_{\theta'}}(\omega),
\end{multline}
with $\theta'-\theta=0$ for homodyne detection and $\theta-\theta' \in \lbrace 0, \pm \pi/2\rbrace$ for heterodyne detection.

Let us now identify the role of each term in Eq.(\ref{eq:inoutPSD}), in analogy to the main article. The first term corresponds to the photodetector shot noise that originates from the correlation function of the interacting mode $\langle\lbrace \hat{a}_\text{in}(\omega),\hat{a}_\text{in}^\dagger(\omega')\rbrace \rangle =\delta(\omega+\omega')/(2\pi)$ when all modes are in vacuum in the displaced frame. We obtain
\begin{equation}
    S^\text{in}_{X_\theta X_{\theta'}}(\omega)=\frac{\cos(\theta-\theta')}{2\pi}.
\end{equation}
The second term corresponds to the interferometric signal of the center-of-mass position which, using Eq.(\ref{eq:spectrumposition}) reads
\begin{equation}
    S_{zz}(\omega)=\frac{4\Gamma}{z_\text{zp}^2}\sin\theta \sin\theta'|\chi(\omega)|^2(S_{FF}^\text{rp}+S_{FF}^\text{th}),
\end{equation}
with
\begin{align}
    S_{FF}^\text{rp}&=\frac{\hbar^2 \Gamma}{2 \pi z_\text{zp}^2},\\
    S_{FF}^\text{th}&=\frac{\hbar^2 \gamma}{4\pi z_\text{zp}^2},
\end{align}
which results from the correlation functions of the driving terms Eq.(\ref{eq:corr1}), and Eq.(\ref{eq:corr2}) and where we have assumed $\hbar \Omega_0 \ll k_B T$. The last two terms are responsible for the sideband asymmetry and result from the correlations between the signal and the interacting mode in vacuum, that is  $\langle \lbrace \hat{z}(\omega), \hat{a}_\text{in}(\omega') \rbrace \rangle = \hbar \sqrt{\Gamma} \chi(\omega') \delta(\omega+\omega')/(2\pi z_\text{zp}).$ They read
\begin{multline}
    S_{X_\theta z}^\text{in}(\omega)+S_{z X_{\theta'}}^\text{in}(\omega)=\frac{\hbar \Gamma}{\pi z_\text{zp}^2}\\
    \times [\chi^*(\omega)\cos\theta\sin\theta'+\chi(\omega) \sin\theta \cos\theta'].
\end{multline}
Combining all of the above results it immediately follows that $S_\text{het}^\text{tot}=S_{PP}/(\bar{P}_\text{ref} \hbar \omega_0)$ and $S_\text{hom}^\text{tot}=S^\text{hom}_{PP}/(\bar{P}_\text{ref} \hbar \omega_0)$ in full agreement with Eq.(\ref{sidebandfinal}) and Eq.(\ref{totpsd3}) in the main article.


\clearpage

\newpage


\begin{thebibliography}{40}%
\makeatletter
\providecommand \@ifxundefined [1]{%
 \@ifx{#1\undefined}
}%
\providecommand \@ifnum [1]{%
 \ifnum #1\expandafter \@firstoftwo
 \else \expandafter \@secondoftwo
 \fi
}%
\providecommand \@ifx [1]{%
 \ifx #1\expandafter \@firstoftwo
 \else \expandafter \@secondoftwo
 \fi
}%
\providecommand \natexlab [1]{#1}%
\providecommand \enquote  [1]{``#1''}%
\providecommand \bibnamefont  [1]{#1}%
\providecommand \bibfnamefont [1]{#1}%
\providecommand \citenamefont [1]{#1}%
\providecommand \href@noop [0]{\@secondoftwo}%
\providecommand \href [0]{\begingroup \@sanitize@url \@href}%
\providecommand \@href[1]{\@@startlink{#1}\@@href}%
\providecommand \@@href[1]{\endgroup#1\@@endlink}%
\providecommand \@sanitize@url [0]{\catcode `\\12\catcode `\$12\catcode
  `\&12\catcode `\#12\catcode `\^12\catcode `\_12\catcode `\%12\relax}%
\providecommand \@@startlink[1]{}%
\providecommand \@@endlink[0]{}%
\providecommand \url  [0]{\begingroup\@sanitize@url \@url }%
\providecommand \@url [1]{\endgroup\@href {#1}{\urlprefix }}%
\providecommand \urlprefix  [0]{URL }%
\providecommand \Eprint [0]{\href }%
\providecommand \doibase [0]{http://dx.doi.org/}%
\providecommand \selectlanguage [0]{\@gobble}%
\providecommand \bibinfo  [0]{\@secondoftwo}%
\providecommand \bibfield  [0]{\@secondoftwo}%
\providecommand \translation [1]{[#1]}%
\providecommand \BibitemOpen [0]{}%
\providecommand \bibitemStop [0]{}%
\providecommand \bibitemNoStop [0]{.\EOS\space}%
\providecommand \EOS [0]{\spacefactor3000\relax}%
\providecommand \BibitemShut  [1]{\csname bibitem#1\endcsname}%
\let\auto@bib@innerbib\@empty
\bibitem [{\citenamefont {Raman}\ and\ \citenamefont
  {Krishnan}(1928)}]{raman28}%
  \BibitemOpen
  \bibfield  {author} {\bibinfo {author} {\bibfnamefont {C.~V.}\ \bibnamefont
  {Raman}}\ and\ \bibinfo {author} {\bibfnamefont {K.~S.}\ \bibnamefont
  {Krishnan}},\ }\href@noop {} {\bibfield  {journal} {\bibinfo  {journal}
  {Nature}\ }\textbf {\bibinfo {volume} {121}},\ \bibinfo {pages} {501}
  (\bibinfo {year} {1928})}\BibitemShut {NoStop}%
\bibitem [{\citenamefont {Placzek}(1934)}]{placzek34}%
  \BibitemOpen
  \bibfield  {author} {\bibinfo {author} {\bibfnamefont {G.}~\bibnamefont
  {Placzek}},\ }\href@noop {} {\emph {\bibinfo {title} {Rayleigh-{S}treuung und
  {R}aman-{E}ffekt}}},\ \bibinfo {series} {Handbuch der {R}adiologie (in
  {G}erman)}, Vol.~\bibinfo {volume} {74}\ (\bibinfo  {publisher} {Leipzig:
  Akademische Verlagsgesellschaft},\ \bibinfo {year} {1934})\ p.\ \bibinfo
  {pages} {209}\BibitemShut {NoStop}%
\bibitem [{\citenamefont {Kip}\ and\ \citenamefont {Meier}(1990)}]{kip90}%
  \BibitemOpen
  \bibfield  {author} {\bibinfo {author} {\bibfnamefont {B.~J.}\ \bibnamefont
  {Kip}}\ and\ \bibinfo {author} {\bibfnamefont {R.~J.}\ \bibnamefont
  {Meier}},\ }\href@noop {} {\bibfield  {journal} {\bibinfo  {journal} {Appl.
  Spectrosc.}\ }\textbf {\bibinfo {volume} {44}},\ \bibinfo {pages} {707}
  (\bibinfo {year} {1990})}\BibitemShut {NoStop}%
\bibitem [{\citenamefont {Wehrmeyer}\ \emph {et~al.}(1996)\citenamefont
  {Wehrmeyer}, \citenamefont {Yeralan},\ and\ \citenamefont
  {Tecu}}]{wehrmeyer96}%
  \BibitemOpen
  \bibfield  {author} {\bibinfo {author} {\bibfnamefont {J.~A.}\ \bibnamefont
  {Wehrmeyer}}, \bibinfo {author} {\bibfnamefont {S.}~\bibnamefont {Yeralan}},
  \ and\ \bibinfo {author} {\bibfnamefont {K.~S.}\ \bibnamefont {Tecu}},\
  }\href@noop {} {\bibfield  {journal} {\bibinfo  {journal} {Appl. Phys. B}\
  }\textbf {\bibinfo {volume} {62}},\ \bibinfo {pages} {21} (\bibinfo {year}
  {1996})}\BibitemShut {NoStop}%
\bibitem [{\citenamefont {Monroe}\ \emph {et~al.}(1995)\citenamefont {Monroe},
  \citenamefont {Meekhof}, \citenamefont {King}, \citenamefont {Jefferts},
  \citenamefont {Itano}, \citenamefont {Wineland},\ and\ \citenamefont
  {Gould}}]{monroe95}%
  \BibitemOpen
  \bibfield  {author} {\bibinfo {author} {\bibfnamefont {C.}~\bibnamefont
  {Monroe}}, \bibinfo {author} {\bibfnamefont {D.~M.}\ \bibnamefont {Meekhof}},
  \bibinfo {author} {\bibfnamefont {B.~E.}\ \bibnamefont {King}}, \bibinfo
  {author} {\bibfnamefont {S.~R.}\ \bibnamefont {Jefferts}}, \bibinfo {author}
  {\bibfnamefont {W.~M.}\ \bibnamefont {Itano}}, \bibinfo {author}
  {\bibfnamefont {D.~J.}\ \bibnamefont {Wineland}}, \ and\ \bibinfo {author}
  {\bibfnamefont {P.}~\bibnamefont {Gould}},\ }\href@noop {} {\bibfield
  {journal} {\bibinfo  {journal} {Phys. Rev. Lett.}\ }\textbf {\bibinfo
  {volume} {75}},\ \bibinfo {pages} {4011} (\bibinfo {year}
  {1995})}\BibitemShut {NoStop}%
\bibitem [{\citenamefont {Brahms}\ \emph {et~al.}(2012)\citenamefont {Brahms},
  \citenamefont {Botter}, \citenamefont {Schreppler}, \citenamefont {Brooks},\
  and\ \citenamefont {Stamper-Kurn}}]{brahms12}%
  \BibitemOpen
  \bibfield  {author} {\bibinfo {author} {\bibfnamefont {N.}~\bibnamefont
  {Brahms}}, \bibinfo {author} {\bibfnamefont {T.}~\bibnamefont {Botter}},
  \bibinfo {author} {\bibfnamefont {S.}~\bibnamefont {Schreppler}}, \bibinfo
  {author} {\bibfnamefont {D.}~\bibnamefont {Brooks}}, \ and\ \bibinfo {author}
  {\bibfnamefont {D.}~\bibnamefont {Stamper-Kurn}},\ }\href@noop {} {\bibfield
  {journal} {\bibinfo  {journal} {Phys. Rev. Lett.}\ }\textbf {\bibinfo
  {volume} {108}},\ \bibinfo {pages} {133601} (\bibinfo {year}
  {2012})}\BibitemShut {NoStop}%
\bibitem [{\citenamefont {Kohler}\ \emph {et~al.}(2017)\citenamefont {Kohler},
  \citenamefont {Spethmann}, \citenamefont {Schreppler},\ and\ \citenamefont
  {Stamper-Kurn}}]{kohler17}%
  \BibitemOpen
  \bibfield  {author} {\bibinfo {author} {\bibfnamefont {J.}~\bibnamefont
  {Kohler}}, \bibinfo {author} {\bibfnamefont {N.}~\bibnamefont {Spethmann}},
  \bibinfo {author} {\bibfnamefont {S.}~\bibnamefont {Schreppler}}, \ and\
  \bibinfo {author} {\bibfnamefont {D.~M.}\ \bibnamefont {Stamper-Kurn}},\
  }\href@noop {} {\bibfield  {journal} {\bibinfo  {journal} {Phys. Rev. Lett.}\
  }\textbf {\bibinfo {volume} {118}},\ \bibinfo {pages} {063604} (\bibinfo
  {year} {2017})}\BibitemShut {NoStop}%
\bibitem [{\citenamefont {Safavi-Naeini}\ \emph {et~al.}(2012)\citenamefont
  {Safavi-Naeini}, \citenamefont {Chan}, \citenamefont {Hill}, \citenamefont
  {Alegre}, \citenamefont {Krause},\ and\ \citenamefont
  {Painter}}]{safavinaeini12}%
  \BibitemOpen
  \bibfield  {author} {\bibinfo {author} {\bibfnamefont {A.~H.}\ \bibnamefont
  {Safavi-Naeini}}, \bibinfo {author} {\bibfnamefont {J.}~\bibnamefont {Chan}},
  \bibinfo {author} {\bibfnamefont {J.~T.}\ \bibnamefont {Hill}}, \bibinfo
  {author} {\bibfnamefont {T.~P.~M.}\ \bibnamefont {Alegre}}, \bibinfo {author}
  {\bibfnamefont {A.}~\bibnamefont {Krause}}, \ and\ \bibinfo {author}
  {\bibfnamefont {O.}~\bibnamefont {Painter}},\ }\href@noop {} {\bibfield
  {journal} {\bibinfo  {journal} {Phys. Rev. Lett.}\ }\textbf {\bibinfo
  {volume} {108}},\ \bibinfo {pages} {033602} (\bibinfo {year}
  {2012})}\BibitemShut {NoStop}%
\bibitem [{\citenamefont {Weinstein}\ \emph {et~al.}(2014)\citenamefont
  {Weinstein}, \citenamefont {Lei}, \citenamefont {Wollman}, \citenamefont
  {Suh}, \citenamefont {Metelmann}, \citenamefont {Clerk},\ and\ \citenamefont
  {Schwab}}]{weinstein14}%
  \BibitemOpen
  \bibfield  {author} {\bibinfo {author} {\bibfnamefont {A.~J.}\ \bibnamefont
  {Weinstein}}, \bibinfo {author} {\bibfnamefont {C.~U.}\ \bibnamefont {Lei}},
  \bibinfo {author} {\bibfnamefont {E.~E.}\ \bibnamefont {Wollman}}, \bibinfo
  {author} {\bibfnamefont {J.}~\bibnamefont {Suh}}, \bibinfo {author}
  {\bibfnamefont {A.}~\bibnamefont {Metelmann}}, \bibinfo {author}
  {\bibfnamefont {A.~A.}\ \bibnamefont {Clerk}}, \ and\ \bibinfo {author}
  {\bibfnamefont {K.~C.}\ \bibnamefont {Schwab}},\ }\href@noop {} {\bibfield
  {journal} {\bibinfo  {journal} {Phys. Rev. X}\ }\textbf {\bibinfo {volume}
  {4}},\ \bibinfo {pages} {041003} (\bibinfo {year} {2014})}\BibitemShut
  {NoStop}%
\bibitem [{\citenamefont {Underwood}\ \emph {et~al.}(2015)\citenamefont
  {Underwood}, \citenamefont {Mason}, \citenamefont {Lee}, \citenamefont {Xu},
  \citenamefont {Jiang}, \citenamefont {Shkarin}, \citenamefont {B{\o}rkje},
  \citenamefont {Girvin},\ and\ \citenamefont {Harris}}]{underwood15}%
  \BibitemOpen
  \bibfield  {author} {\bibinfo {author} {\bibfnamefont {M.}~\bibnamefont
  {Underwood}}, \bibinfo {author} {\bibfnamefont {D.}~\bibnamefont {Mason}},
  \bibinfo {author} {\bibfnamefont {D.}~\bibnamefont {Lee}}, \bibinfo {author}
  {\bibfnamefont {H.}~\bibnamefont {Xu}}, \bibinfo {author} {\bibfnamefont
  {L.}~\bibnamefont {Jiang}}, \bibinfo {author} {\bibfnamefont {A.~B.}\
  \bibnamefont {Shkarin}}, \bibinfo {author} {\bibfnamefont {K.}~\bibnamefont
  {B{\o}rkje}}, \bibinfo {author} {\bibfnamefont {S.~M.}\ \bibnamefont
  {Girvin}}, \ and\ \bibinfo {author} {\bibfnamefont {J.~G.~E.}\ \bibnamefont
  {Harris}},\ }\href@noop {} {\bibfield  {journal} {\bibinfo  {journal} {Phys.
  Rev. A}\ }\textbf {\bibinfo {volume} {92}},\ \bibinfo {pages} {061801(R)}
  (\bibinfo {year} {2015})}\BibitemShut {NoStop}%
\bibitem [{\citenamefont {Purdy}\ \emph {et~al.}(2015)\citenamefont {Purdy},
  \citenamefont {Yu}, \citenamefont {Kampel}, \citenamefont {Peterson},
  \citenamefont {Cicak}, \citenamefont {Simmonds},\ and\ \citenamefont
  {Regal}}]{purdy15}%
  \BibitemOpen
  \bibfield  {author} {\bibinfo {author} {\bibfnamefont {T.~P.}\ \bibnamefont
  {Purdy}}, \bibinfo {author} {\bibfnamefont {P.-L.}\ \bibnamefont {Yu}},
  \bibinfo {author} {\bibfnamefont {N.~S.}\ \bibnamefont {Kampel}}, \bibinfo
  {author} {\bibfnamefont {R.~W.}\ \bibnamefont {Peterson}}, \bibinfo {author}
  {\bibfnamefont {K.}~\bibnamefont {Cicak}}, \bibinfo {author} {\bibfnamefont
  {R.~W.}\ \bibnamefont {Simmonds}}, \ and\ \bibinfo {author} {\bibfnamefont
  {C.~A.}\ \bibnamefont {Regal}},\ }\href@noop {} {\bibfield  {journal}
  {\bibinfo  {journal} {Phys. Rev. A}\ }\textbf {\bibinfo {volume} {92}},\
  \bibinfo {pages} {031802(R)} (\bibinfo {year} {2015})}\BibitemShut {NoStop}%
\bibitem [{\citenamefont {Sudhir}\ \emph {et~al.}(2017)\citenamefont {Sudhir},
  \citenamefont {Wilson}, \citenamefont {Schilling}, \citenamefont {Sch\"utz},
  \citenamefont {Fedorov}, \citenamefont {Ghadimi}, \citenamefont
  {Nunnenkamp},\ and\ \citenamefont {Kippenberg}}]{sudhir17}%
  \BibitemOpen
  \bibfield  {author} {\bibinfo {author} {\bibfnamefont {V.}~\bibnamefont
  {Sudhir}}, \bibinfo {author} {\bibfnamefont {D.~J.}\ \bibnamefont {Wilson}},
  \bibinfo {author} {\bibfnamefont {R.}~\bibnamefont {Schilling}}, \bibinfo
  {author} {\bibfnamefont {H.}~\bibnamefont {Sch\"utz}}, \bibinfo {author}
  {\bibfnamefont {S.~A.}\ \bibnamefont {Fedorov}}, \bibinfo {author}
  {\bibfnamefont {A.~H.}\ \bibnamefont {Ghadimi}}, \bibinfo {author}
  {\bibfnamefont {A.}~\bibnamefont {Nunnenkamp}}, \ and\ \bibinfo {author}
  {\bibfnamefont {T.~J.}\ \bibnamefont {Kippenberg}},\ }\href@noop {}
  {\bibfield  {journal} {\bibinfo  {journal} {Phys. Rev. X}\ }\textbf {\bibinfo
  {volume} {7}},\ \bibinfo {pages} {011001} (\bibinfo {year}
  {2017})}\BibitemShut {NoStop}%
\bibitem [{\citenamefont {Tebbenjohanns}\ \emph {et~al.}(2020)\citenamefont
  {Tebbenjohanns}, \citenamefont {Frimmer}, \citenamefont {Jain}, \citenamefont
  {Windey},\ and\ \citenamefont {Novotny}}]{tebbenjohanns20}%
  \BibitemOpen
  \bibfield  {author} {\bibinfo {author} {\bibfnamefont {F.}~\bibnamefont
  {Tebbenjohanns}}, \bibinfo {author} {\bibfnamefont {M.}~\bibnamefont
  {Frimmer}}, \bibinfo {author} {\bibfnamefont {V.}~\bibnamefont {Jain}},
  \bibinfo {author} {\bibfnamefont {D.}~\bibnamefont {Windey}}, \ and\ \bibinfo
  {author} {\bibfnamefont {L.}~\bibnamefont {Novotny}},\ }\href@noop {}
  {\bibfield  {journal} {\bibinfo  {journal} {Phys. Rev. Lett.}\ }\textbf
  {\bibinfo {volume} {124}},\ \bibinfo {pages} {013603} (\bibinfo {year}
  {2020})}\BibitemShut {NoStop}%
\bibitem [{\citenamefont {Khalili}\ \emph {et~al.}(2012)\citenamefont
  {Khalili}, \citenamefont {Miao}, \citenamefont {Yang}, \citenamefont
  {Safavi-Naeini}, \citenamefont {Painter},\ and\ \citenamefont
  {Chen}}]{khalili12}%
  \BibitemOpen
  \bibfield  {author} {\bibinfo {author} {\bibfnamefont {F.~Y.}\ \bibnamefont
  {Khalili}}, \bibinfo {author} {\bibfnamefont {H.}~\bibnamefont {Miao}},
  \bibinfo {author} {\bibfnamefont {H.}~\bibnamefont {Yang}}, \bibinfo {author}
  {\bibfnamefont {A.~H.}\ \bibnamefont {Safavi-Naeini}}, \bibinfo {author}
  {\bibfnamefont {O.}~\bibnamefont {Painter}}, \ and\ \bibinfo {author}
  {\bibfnamefont {Y.}~\bibnamefont {Chen}},\ }\href@noop {} {\bibfield
  {journal} {\bibinfo  {journal} {Phys. Rev. A}\ }\textbf {\bibinfo {volume}
  {86}},\ \bibinfo {pages} {033840} (\bibinfo {year} {2012})}\BibitemShut
  {NoStop}%
\bibitem [{\citenamefont {B{\o}rkje}(2016)}]{borkje16}%
  \BibitemOpen
  \bibfield  {author} {\bibinfo {author} {\bibfnamefont {K.}~\bibnamefont
  {B{\o}rkje}},\ }\href@noop {} {\bibfield  {journal} {\bibinfo  {journal}
  {Phys. Rev. A}\ }\textbf {\bibinfo {volume} {94}},\ \bibinfo {pages} {043816}
  (\bibinfo {year} {2016})}\BibitemShut {NoStop}%
\bibitem [{\citenamefont {Machado}\ and\ \citenamefont
  {Blanter}(2021)}]{machado21}%
  \BibitemOpen
  \bibfield  {author} {\bibinfo {author} {\bibfnamefont {J.~D.~P.}\
  \bibnamefont {Machado}}\ and\ \bibinfo {author} {\bibfnamefont {Y.~M.}\
  \bibnamefont {Blanter}},\ }\href@noop {} {\bibfield  {journal} {\bibinfo
  {journal} {{SciPost Phys. Core}}\ } \textbf {\bibinfo {volume} {{5}}}, \ \bibinfo {pages} {{034}}
  (\bibinfo {year} {{2022}})}\BibitemShut{NoStop}%
\bibitem [{\citenamefont {Marshall}(1963)}]{marshall63}%
  \BibitemOpen
  \bibfield  {author} {\bibinfo {author} {\bibfnamefont {T.~W.}\ \bibnamefont
  {Marshall}},\ }\href@noop {} {\bibfield  {journal} {\bibinfo  {journal}
  {Proc. R. Soc. A}\ }\textbf {\bibinfo {volume} {276}},\ \bibinfo {pages}
  {475} (\bibinfo {year} {1963})}\BibitemShut {NoStop}%
\bibitem [{\citenamefont {Boyer}(1975{\natexlab{a}})}]{boyer75a}%
  \BibitemOpen
  \bibfield  {author} {\bibinfo {author} {\bibfnamefont {T.~H.}\ \bibnamefont
  {Boyer}},\ }\href@noop {} {\bibfield  {journal} {\bibinfo  {journal} {Phys.
  Rev. D}\ }\textbf {\bibinfo {volume} {11}},\ \bibinfo {pages} {790} (\bibinfo
  {year} {1975}{\natexlab{a}})}\BibitemShut {NoStop}%
\bibitem [{\citenamefont {Boyer}(2011)}]{boyer11}%
  \BibitemOpen
  \bibfield  {author} {\bibinfo {author} {\bibfnamefont {T.~H.}\ \bibnamefont
  {Boyer}},\ }\href@noop {} {\bibfield  {journal} {\bibinfo  {journal} {Am. J.
  Phys.}\ }\textbf {\bibinfo {volume} {79}},\ \bibinfo {pages} {1163} (\bibinfo
  {year} {2011})}\BibitemShut {NoStop}%
\bibitem [{\citenamefont {Novotny}\ and\ \citenamefont
  {Hecht}(2012)}]{novotny12}%
  \BibitemOpen
  \bibfield  {author} {\bibinfo {author} {\bibfnamefont {L.}~\bibnamefont
  {Novotny}}\ and\ \bibinfo {author} {\bibfnamefont {B.}~\bibnamefont
  {Hecht}},\ }\href@noop {} {\emph {\bibinfo {title} {Principles of
  Nano-Optics}}},\ \bibinfo {edition} {2nd}\ ed.\ (\bibinfo  {publisher}
  {Cambridge University Press},\ \bibinfo {address} {Cambridge},\ \bibinfo
  {year} {2012})\ Chap.~\bibinfo {chapter} {15}\BibitemShut {NoStop}%
\bibitem [{\citenamefont {Callen}\ and\ \citenamefont
  {Welton}(1951)}]{callen51}%
  \BibitemOpen
  \bibfield  {author} {\bibinfo {author} {\bibfnamefont {H.~B.}\ \bibnamefont
  {Callen}}\ and\ \bibinfo {author} {\bibfnamefont {T.~A.}\ \bibnamefont
  {Welton}},\ }\href@noop {} {\bibfield  {journal} {\bibinfo  {journal} {Phys.
  Rev.}\ }\textbf {\bibinfo {volume} {83}},\ \bibinfo {pages} {34} (\bibinfo
  {year} {1951})}\BibitemShut {NoStop}%
\bibitem [{\citenamefont {Agarwal}(1975)}]{agarwal75}%
  \BibitemOpen
  \bibfield  {author} {\bibinfo {author} {\bibfnamefont {G.~S.}\ \bibnamefont
  {Agarwal}},\ }\href@noop {} {\bibfield  {journal} {\bibinfo  {journal} {Phys.
  Rev. A}\ }\textbf {\bibinfo {volume} {11}},\ \bibinfo {pages} {230} (\bibinfo
  {year} {1975})}\BibitemShut {NoStop}%
\bibitem [{\citenamefont {Rytov}\ \emph {et~al.}(1987)\citenamefont {Rytov},
  \citenamefont {Kravtsov},\ and\ \citenamefont {Tatarskii}}]{rytov88}%
  \BibitemOpen
  \bibfield  {author} {\bibinfo {author} {\bibfnamefont {S.~M.}\ \bibnamefont
  {Rytov}}, \bibinfo {author} {\bibfnamefont {Y.~A.}\ \bibnamefont {Kravtsov}},
  \ and\ \bibinfo {author} {\bibfnamefont {V.~I.}\ \bibnamefont {Tatarskii}},\
  }in\ \href@noop {} {\emph {\bibinfo {booktitle} {Principles of Statistical
  Radiophysics}}},\ Vol.~\bibinfo {volume} {3}\ (\bibinfo  {publisher}
  {Springer-Verlag},\ \bibinfo {address} {Berlin},\ \bibinfo {year} {1987})\
  Chap.~\bibinfo {chapter} {3}, pp.\ \bibinfo {pages} {109--173}\BibitemShut
  {NoStop}%
\bibitem [{\citenamefont {Eckhardt}(1982)}]{eckhardt82}%
  \BibitemOpen
  \bibfield  {author} {\bibinfo {author} {\bibfnamefont {W.}~\bibnamefont
  {Eckhardt}},\ }\href@noop {} {\bibfield  {journal} {\bibinfo  {journal} {Opt.
  Commun.}\ }\textbf {\bibinfo {volume} {41}},\ \bibinfo {pages} {305}
  (\bibinfo {year} {1982})}\BibitemShut {NoStop}%
\bibitem [{\citenamefont {Boyer}(1969)}]{boyer69}%
  \BibitemOpen
  \bibfield  {author} {\bibinfo {author} {\bibfnamefont {T.~H.}\ \bibnamefont
  {Boyer}},\ }\href@noop {} {\bibfield  {journal} {\bibinfo  {journal} {Phys.
  Rev.}\ }\textbf {\bibinfo {volume} {182}},\ \bibinfo {pages} {1374} (\bibinfo
  {year} {1969})}\BibitemShut {NoStop}%
\bibitem [{\citenamefont {Boyer}(1975{\natexlab{b}})}]{boyer75b}%
  \BibitemOpen
  \bibfield  {author} {\bibinfo {author} {\bibfnamefont {T.~H.}\ \bibnamefont
  {Boyer}},\ }\href@noop {} {\bibfield  {journal} {\bibinfo  {journal} {Phys.
  Rev. D}\ }\textbf {\bibinfo {volume} {11}},\ \bibinfo {pages} {809} (\bibinfo
  {year} {1975}{\natexlab{b}})}\BibitemShut {NoStop}%
\bibitem [{\citenamefont {Henkel}\ \emph {et~al.}(2002)\citenamefont {Henkel},
  \citenamefont {Joulain}, \citenamefont {Mulet},\ and\ \citenamefont
  {Greffet}}]{henkel02}%
  \BibitemOpen
  \bibfield  {author} {\bibinfo {author} {\bibfnamefont {C.}~\bibnamefont
  {Henkel}}, \bibinfo {author} {\bibfnamefont {K.}~\bibnamefont {Joulain}},
  \bibinfo {author} {\bibfnamefont {J.-P.}\ \bibnamefont {Mulet}}, \ and\
  \bibinfo {author} {\bibfnamefont {J.-J.}\ \bibnamefont {Greffet}},\
  }\href@noop {} {\bibfield  {journal} {\bibinfo  {journal} {J. Opt. A: Pure
  Appl. Opt.}\ }\textbf {\bibinfo {volume} {4}},\ \bibinfo {pages} {S109}
  (\bibinfo {year} {2002})}\BibitemShut {NoStop}%
\bibitem [{\citenamefont {Mulet}\ \emph {et~al.}(2001)\citenamefont {Mulet},
  \citenamefont {Joulain}, \citenamefont {Carminati},\ and\ \citenamefont
  {Greffet}}]{mulet01}%
  \BibitemOpen
  \bibfield  {author} {\bibinfo {author} {\bibfnamefont {J.~P.}\ \bibnamefont
  {Mulet}}, \bibinfo {author} {\bibfnamefont {K.}~\bibnamefont {Joulain}},
  \bibinfo {author} {\bibfnamefont {R.}~\bibnamefont {Carminati}}, \ and\
  \bibinfo {author} {\bibfnamefont {J.~J.}\ \bibnamefont {Greffet}},\
  }\href@noop {} {\bibfield  {journal} {\bibinfo  {journal} {Appl. Phys.
  Lett.}\ }\textbf {\bibinfo {volume} {78}},\ \bibinfo {pages} {2931} (\bibinfo
  {year} {2001})}\BibitemShut {NoStop}%
\bibitem [{\citenamefont {Cole}\ and\ \citenamefont {Zou}(2003)}]{cole03}%
  \BibitemOpen
  \bibfield  {author} {\bibinfo {author} {\bibfnamefont {D.~C.}\ \bibnamefont
  {Cole}}\ and\ \bibinfo {author} {\bibfnamefont {Y.}~\bibnamefont {Zou}},\
  }\href@noop {} {\bibfield  {journal} {\bibinfo  {journal} {Phys. Lett. A}\
  }\textbf {\bibinfo {volume} {317}},\ \bibinfo {pages} {14} (\bibinfo {year}
  {2003})}\BibitemShut {NoStop}%
\bibitem [{\citenamefont {Zurita-S\'anchez}\ \emph {et~al.}(2004)\citenamefont
  {Zurita-S\'anchez}, \citenamefont {Greffet},\ and\ \citenamefont
  {Novotny}}]{zurita04}%
  \BibitemOpen
  \bibfield  {author} {\bibinfo {author} {\bibfnamefont {J.~R.}\ \bibnamefont
  {Zurita-S\'anchez}}, \bibinfo {author} {\bibfnamefont {J.-J.}\ \bibnamefont
  {Greffet}}, \ and\ \bibinfo {author} {\bibfnamefont {L.}~\bibnamefont
  {Novotny}},\ }\href {\doibase 10.1103/PhysRevA.69.022902} {\bibfield
  {journal} {\bibinfo  {journal} {Phys. Rev. A}\ }\textbf {\bibinfo {volume}
  {69}},\ \bibinfo {pages} {022902} (\bibinfo {year} {2004})}\BibitemShut
  {NoStop}%
\bibitem [{\citenamefont {Schottky}(1918)}]{schottky18}%
  \BibitemOpen
  \bibfield  {author} {\bibinfo {author} {\bibfnamefont {W.}~\bibnamefont
  {Schottky}},\ }\href@noop {} {\bibfield  {journal} {\bibinfo  {journal} {Ann.
  d. Phys.}\ }\textbf {\bibinfo {volume} {362}},\ \bibinfo {pages} {541}
  (\bibinfo {year} {1918})}\BibitemShut {NoStop}%
\bibitem [{\citenamefont {Millen}\ \emph {et~al.}(2020)\citenamefont {Millen},
  \citenamefont {Monteiro}, \citenamefont {Pettit},\ and\ \citenamefont
  {Vamivakas}}]{millen20}%
  \BibitemOpen
  \bibfield  {author} {\bibinfo {author} {\bibfnamefont {J.}~\bibnamefont
  {Millen}}, \bibinfo {author} {\bibfnamefont {T.~S.}\ \bibnamefont
  {Monteiro}}, \bibinfo {author} {\bibfnamefont {R.}~\bibnamefont {Pettit}}, \
  and\ \bibinfo {author} {\bibfnamefont {A.~N.}\ \bibnamefont {Vamivakas}},\
  }\href {\doibase 10.1088/1361-6633/ab6100} {\bibfield  {journal} {\bibinfo
  {journal} {Rep. Prog. Phys.}\ }\textbf {\bibinfo {volume} {83}},\ \bibinfo
  {pages} {026401} (\bibinfo {year} {2020})}\BibitemShut {NoStop}%
\bibitem [{\citenamefont {Gonzalez-Ballestero}\ \emph
  {et~al.}(2021)\citenamefont {Gonzalez-Ballestero}, \citenamefont
  {Aspelmeyer}, \citenamefont {Novotny}, \citenamefont {Quidant},\ and\
  \citenamefont {Romero-Isart}}]{gonzales21}%
  \BibitemOpen
  \bibfield  {author} {\bibinfo {author} {\bibfnamefont {C.}~\bibnamefont
  {Gonzalez-Ballestero}}, \bibinfo {author} {\bibfnamefont {M.}~\bibnamefont
  {Aspelmeyer}}, \bibinfo {author} {\bibfnamefont {L.}~\bibnamefont {Novotny}},
  \bibinfo {author} {\bibfnamefont {R.}~\bibnamefont {Quidant}}, \ and\
  \bibinfo {author} {\bibfnamefont {O.}~\bibnamefont {Romero-Isart}},\ }\href
  {\doibase 10.1126/science.abg3027} {\bibfield  {journal} {\bibinfo  {journal}
  {Science}\ }\textbf {\bibinfo {volume} {374}},\ \bibinfo {pages} {eabg3027}
  (\bibinfo {year} {2021})}\BibitemShut {NoStop}%
\bibitem [{\citenamefont {Pirandola}\ \emph {et~al.}(2003)\citenamefont
  {Pirandola}, \citenamefont {Mancini}, \citenamefont {Vitali},\ and\
  \citenamefont {Tombesi}}]{pirandola03}%
  \BibitemOpen
  \bibfield  {author} {\bibinfo {author} {\bibfnamefont {S.}~\bibnamefont
  {Pirandola}}, \bibinfo {author} {\bibfnamefont {S.}~\bibnamefont {Mancini}},
  \bibinfo {author} {\bibfnamefont {D.}~\bibnamefont {Vitali}}, \ and\ \bibinfo
  {author} {\bibfnamefont {P.}~\bibnamefont {Tombesi}},\ }\href {\doibase
  10.1103/PhysRevA.68.062317} {\bibfield  {journal} {\bibinfo  {journal} {Phys.
  Rev. A}\ }\textbf {\bibinfo {volume} {68}},\ \bibinfo {pages} {062317}
  (\bibinfo {year} {2003})}\BibitemShut {NoStop}%
\bibitem [{\citenamefont {{Maurer {\em et al.}}}(2022)}]{maurer22}%
  \BibitemOpen
  \bibfield  {author} {\bibinfo {author} {\bibfnamefont {P.}~\bibnamefont
  {{Maurer {\em et al.}}}},\ }\href@noop {} {\bibfield  {journal} {\bibinfo
  {journal} {in preparation}\ } (\bibinfo {year} {2022})}\BibitemShut {NoStop}%
\bibitem [{\citenamefont {Gardiner}\ and\ \citenamefont
  {Zoller}(2004)}]{gardiner2004}%
  \BibitemOpen
  \bibfield  {author} {\bibinfo {author} {\bibfnamefont {C.~W.}\ \bibnamefont
  {Gardiner}}\ and\ \bibinfo {author} {\bibfnamefont {P.}~\bibnamefont
  {Zoller}},\ }\href@noop {} {\emph {\bibinfo {title} {Quantum noise}}},\
  \bibinfo {edition} {3rd}\ ed.\ (\bibinfo  {publisher} {Springer},\ \bibinfo
  {address} {Berlin, Germany},\ \bibinfo {year} {2004})\BibitemShut {NoStop}%
\bibitem [{\citenamefont {Militaru}\ \emph {et~al.}(2022)\citenamefont
  {Militaru}, \citenamefont {Rossi}, \citenamefont {Tebbenjohanns},
  \citenamefont {Romero-Isart}, \citenamefont {Frimmer},\ and\ \citenamefont
  {Novotny}}]{militaru2022}%
  \BibitemOpen
  \bibfield  {author} {\bibinfo {author} {\bibfnamefont {A.}~\bibnamefont
  {Militaru}}, \bibinfo {author} {\bibfnamefont {M.}~\bibnamefont {Rossi}},
  \bibinfo {author} {\bibfnamefont {F.}~\bibnamefont {Tebbenjohanns}}, \bibinfo
  {author} {\bibfnamefont {O.}~\bibnamefont {Romero-Isart}}, \bibinfo {author}
  {\bibfnamefont {M.}~\bibnamefont {Frimmer}}, \ and\ \bibinfo {author}
  {\bibfnamefont {L.}~\bibnamefont {Novotny}},\ }\href@noop {} {\bibfield
  {journal} {\bibinfo  {journal} {arXiv:2202.09063}\ } (\bibinfo {year}
  {2022})}\BibitemShut {NoStop}%
\bibitem [{\citenamefont {Gardiner}\ and\ \citenamefont
  {Collett}(1985)}]{gardiner1985}%
  \BibitemOpen
  \bibfield  {author} {\bibinfo {author} {\bibfnamefont {C.~W.}\ \bibnamefont
  {Gardiner}}\ and\ \bibinfo {author} {\bibfnamefont {M.~J.}\ \bibnamefont
  {Collett}},\ }\href {\doibase 10.1103/PhysRevA.31.3761} {\bibfield  {journal}
  {\bibinfo  {journal} {Phys. Rev. A}\ }\textbf {\bibinfo {volume} {31}},\
  \bibinfo {pages} {3761} (\bibinfo {year} {1985})}\BibitemShut {NoStop}%
\bibitem [{\citenamefont {Magrini}\ \emph {et~al.}(2022)\citenamefont
  {Magrini}, \citenamefont {Camarena-ChÃ¡vez}, \citenamefont {Bach},
  \citenamefont {Johnson},\ and\ \citenamefont {Aspelmeyer}}]{magrini2022}%
  \BibitemOpen
  \bibfield  {author} {\bibinfo {author} {\bibfnamefont {L.}~\bibnamefont
  {Magrini}}, \bibinfo {author} {\bibfnamefont {V.~A.}\ \bibnamefont
  {Camarena-ChÃ¡vez}}, \bibinfo {author} {\bibfnamefont {C.}~\bibnamefont
  {Bach}}, \bibinfo {author} {\bibfnamefont {A.}~\bibnamefont {Johnson}}, \
  and\ \bibinfo {author} {\bibfnamefont {M.}~\bibnamefont {Aspelmeyer}},\
  }\href@noop {} {\bibfield  {journal} {\bibinfo  {journal} {arXiv:2202.09322}\
  } (\bibinfo {year} {2022})}\BibitemShut {NoStop}%
\bibitem [{\citenamefont {Tebbenjohanns}\ \emph {et~al.}(2019)\citenamefont
  {Tebbenjohanns}, \citenamefont {Frimmer},\ and\ \citenamefont
  {Novotny}}]{tebbenjohanns2019}%
  \BibitemOpen
  \bibfield  {author} {\bibinfo {author} {\bibfnamefont {F.}~\bibnamefont
  {Tebbenjohanns}}, \bibinfo {author} {\bibfnamefont {M.}~\bibnamefont
  {Frimmer}}, \ and\ \bibinfo {author} {\bibfnamefont {L.}~\bibnamefont
  {Novotny}},\ }\href {\doibase 10.1103/PhysRevA.100.043821} {\bibfield
  {journal} {\bibinfo  {journal} {Phys. Rev. A}\ }\textbf {\bibinfo {volume}
  {100}},\ \bibinfo {pages} {043821} (\bibinfo {year} {2019})}\BibitemShut
  {NoStop}%
\end{thebibliography}
\end{document}